\documentclass[prb,twocolumn,superscriptaddress,aps,amsmath,amsfonts,notitlepage,showpacs]{revtex4-1}
\usepackage{amsmath,amsfonts,amssymb,dsfont,graphicx,psfrag,bm}

\usepackage{color}

\def\red{\color[rgb]{1,0,0}}
\def\blue{\color[rgb]{0,0,1}}

\def\be{\begin{equation}}
\def\ee{\end{equation}}
\def\bea{\begin{eqnarray}}
\def\eea{\end{eqnarray}}
\def\bpm{\begin{pmatrix}}
\def\epm{\end{pmatrix}}

\def\nn{\nonumber}

\def\tr{\mathop{\rm tr}}

\newcommand{\ve}{\varepsilon}

\newcommand{\om}{\omega}

\newcommand{\g}{\gamma}
\newcommand{\D}{\Delta}

\newcommand{\Pn}{\pi}
\newcommand{\Pc}{\pi^\dagger}

\newcommand{\LL}[1]{\left| #1\right\rangle}

\begin{document}
\title{New Dirac points and multiple Landau level crossings in biased trilayer graphene}
\author{Maksym Serbyn} 
\affiliation{Department of Physics, Massachusetts Institute of Technology, Cambridge, MA, USA}
\author{Dmitry A. Abanin}
\affiliation{Perimeter Institute for Theoretical Physical, Waterloo, ON, Canada}
\affiliation{Institute for Quantum Computing, Waterloo, ON, Canada}
\affiliation{Department of Physics, Harvard University, Cambridge, MA, USA}
\date{\today}
\begin{abstract}
Recently a new high-mobility Dirac material, trilayer graphene, was realized experimentally. The band structure of $ABA$-stacked trilayer graphene consists of a monolayer-like and a bilayer-like pairs of bands. Here we study electronic properties of $ABA$-stacked trilayer graphene biased by a perpendicular electric field. We find that the combination of the bias and  trigonal warping gives rise to a set of new Dirac points: in each valley, seven species of Dirac fermions with small masses of order of a few meV emerge. The positions and masses of the emergent Dirac fermions are tunable by bias, and one group of Dirac fermions becomes massless at a certain bias value. Therefore, in contrast to bilayer graphene, the conductivity at the neutrality point is expected to show non-monotonic behavior, becoming of the order of a few $e^2/h$ when some Dirac masses vanish. Further, we analyze the evolution of Landau level spectrum as a function of bias. Emergence of new Dirac points in the band structure translates into new three-fold-degenerate groups of Landau levels. This leads to an anomalous quantum Hall effect, in which some quantum Hall steps have a height of $3e^2/h$. At an intermediate bias, the degeneracies of all Landau levels get lifted, and in this regime all quantum Hall plateaus are spaced by $e^2/h$. Finally, we show that the pattern of Landau level crossings is very sensitive to certain band structure parameters, and can therefore provide a useful tool for determining their precise values.
\end{abstract}
\maketitle

\section{Introduction \label{Sec:Intro}}

Single- and multi-layer graphene materials are gapless 2D semimetals with unusual band structure, in which low-energy excitations have non-trivial Berry's phases.~\cite{CastroNeto09} In particular, low-energy excitations in graphene are two species (valleys) of massless Dirac-like fermions with linear dispersion and Berry's phase $\pi$. The excitations in bilayer graphene are chiral fermions with a parabolic dispersion relation and Berry's phase of $2\pi$.~\cite{CastroNeto09} The unusual nature of excitations underlies  fundamental phenomena found in these materials, including Klein tunneling, weak anti-localization, anomalous quantum Hall effects, as well as novel symmetry-broken states in high magnetic field.~\cite{CastroNeto09}

One important distinct feature of bilayer graphene (as compared to monolayer graphene and other 2D electron systems) is a unique tunability of its band structure. By applying a perpendicular electric field~\cite{Castro07,Oostinga08,Zhang09} which induces an asymmetry between the top and bottom layer, it is possible to induce a band gap of up to $250\, {\rm meV}$.~\cite{Zhang09} Thus, bilayer graphene can be turned into a semiconductor with a widely tunable band gap, which makes it attractive for various device applications, and provides a way to study chiral carriers in new regimes. 

Very recently, a new carbon-based semimetal, trilayer graphene, has been realized experimentally.~\cite{ThitiExp,Lau11,Heinz11,Kumar11} Experimentally studied trilayer graphene samples had a high mobility, and exhibited quantum Hall effect (QHE). Inspired by these experiments, in this paper we study the effect of perpendicular electric field on the electronic properties of trilayer graphene.  First, we aim to understand how the band structure of trilayer graphene transforms under bias, and to explore new possibilities offered by its tunability. Second, we would like to develop an understanding of Landau level (LL) evolution and LL crossings under bias. Given that band structure of trilayer graphene is quite complicated, it is necessary to first understand single-particle effects in LL evolution, before addressing interaction phenomena, including quantum Hall ferromagnetism and fractional QHE. 

Depending on the stacking ($ABA$ or $ABC$), the band structure of trilayer graphene can be very different~\cite{McCannGateABA,McCannABCBerry,Heinz11,MacDonald-ABCband,Aoki-mult}. Throughout the paper, we will focus on the case of Bernal-stacked ($ABA$) trilayer graphene.~\cite{ThitiExp,Lau11} In the absence of bias, band structure of $ABA$-trilayer graphene can be viewed as a combination of independent overlapping monolayer-like (Berry's phase $\pi$) and bilayer-like bands (Berry's phase $2\pi$).~\cite{McCannGateABA,McCannParityValleyMultilayer,Koshino-mult2,Koshino-mult} Both bands are gapped, in contrast to gapless monolayer and bilayer graphene, however, the monolayer-like and bilayer-like gaps are overlapping such that there is no band gap. In contrast, $ABC$-stacked trilayer graphene has a Dirac points with cubic band touching and  Berry's phase $3\pi$,~\cite{McCannABCBerry,MacDonald-ABCband} split by different symmetry-breaking terms. 

Previously, it was noted that perpendicular electric field completely transforms the band structure of trilayer graphene, hybridizing the monolayer-like and bilayer-like bands,~\cite{McCannGateABA,Peeters-EF,*Peeters-mult,*Peeters-mult2} and generally opening only a small band gap. Here we explore the transformation of the band structure in detail. Most interestingly, we find that as a result of trigonal warping, at bias $\sim 0.5$ V/nm,~\footnote{In this estimate we use values of tight binding parameters from Ref.~\onlinecite{ThitiExp} and value $\ve=2$ for the dielectric constant of $ABA$ trilayer graphene.} \emph{seven} new Dirac points (DP) emerge in each valley. One of the DPs is situated at the $K_+$ ($K_-$) point in the Brillouin zone; the remaining six consist of two groups of three, related to each other by $2\pi/3$ rotations. The emergent Dirac fermions are massive, with both their mass and velocity being tunable by an electric field. At electric field $\sim 0.9$ V/nm,~\cite{Note1} the band gap closes for three DPs. For higher bias the mass of these DPs changes sign, and overall gap increases monotonically. 

We analyze the evolution of LLs under bias, finding a very complex pattern of levels crossings and splittings. We illustrate the extreme sensitivity of this pattern with respect to band structure parameters. Even small variations of certain parameters change the pattern of crossings at small bias. Therefore, the Landau level crossings should provide a useful tool for precise determination of the band structure parameters, which are still a subject of debate.~\cite{Dress-tb,ThitiExp}  The transformation of the band structure at high bias leads to an emergent three-fold degeneracy of some Landau levels in magnetic field.  It is worth noting that the evolution of LLs with bias has been previously discussed by McCann and Koshino in Ref.~\onlinecite{McCannParityValleyMultilayer}, who, however, considered an idealized model that neglected several smaller hopping parameters~(see also Ref.~\onlinecite{Peeters-LL}). Here we consider the full tight-binding model,~\cite{ThitiExp} finding that the smaller parameters play an important role in determining the pattern of LL crossings.

Our results also imply that single-particle physics in $ABA$ trilayer graphene is much more complicated as compared to other chiral materials, such as monolayer and bilayer graphene, and it should be understood before addressing the interaction effects. On the positive side, the tunability of the band structure may prove advantageous in studying and tuning correlated states, including fractional quantum Hall states,~\cite{Papic11,Papic12} as discussed in the end of the paper. 

The rest of the paper is organized as follows: in the next Section we introduce the tight-binding description for the $ABA$-stacked trilayer graphene and review decoupling of its band structure into monolayer-like and bilayer-like blocks~\cite{McCannGateABA}. In Section~\ref{Sec:Band} we discuss the evolution of the band structure as a function of perpendicular electric field.  Landau level spectrum as a function of magnetic field and electric field is studied in Section~\ref{Sec:LLs}. Finally, in Section~\ref{Sec:Discussion}, we provide a brief summary of our results and present a broader outlook.

\section{Model and review of unbiased case \label{Sec:model}}

In this Section, we introduce the tight-binding model of the $ABA$-stacked trilayer graphene and briefly review the derivation of low-energy effective Hamiltonian in the absence of bias following Ref.~\onlinecite{McCannGateABA}. We will discuss that in the absence of perpendicular electric field~(also referred to as displacement field), the band structure can be viewed as two independent bands: one bilayer-graphene-like and the other monolayer-graphene-like.~\cite{McCannGateABA,McCannParityValleyMultilayer,Koshino-mult2,Koshino-mult}

\begin{figure}
\begin{center}
\begin{minipage}[t]{0.7\columnwidth}
\vspace{4pt}
\mbox{}\par
\includegraphics[width=0.9\columnwidth]{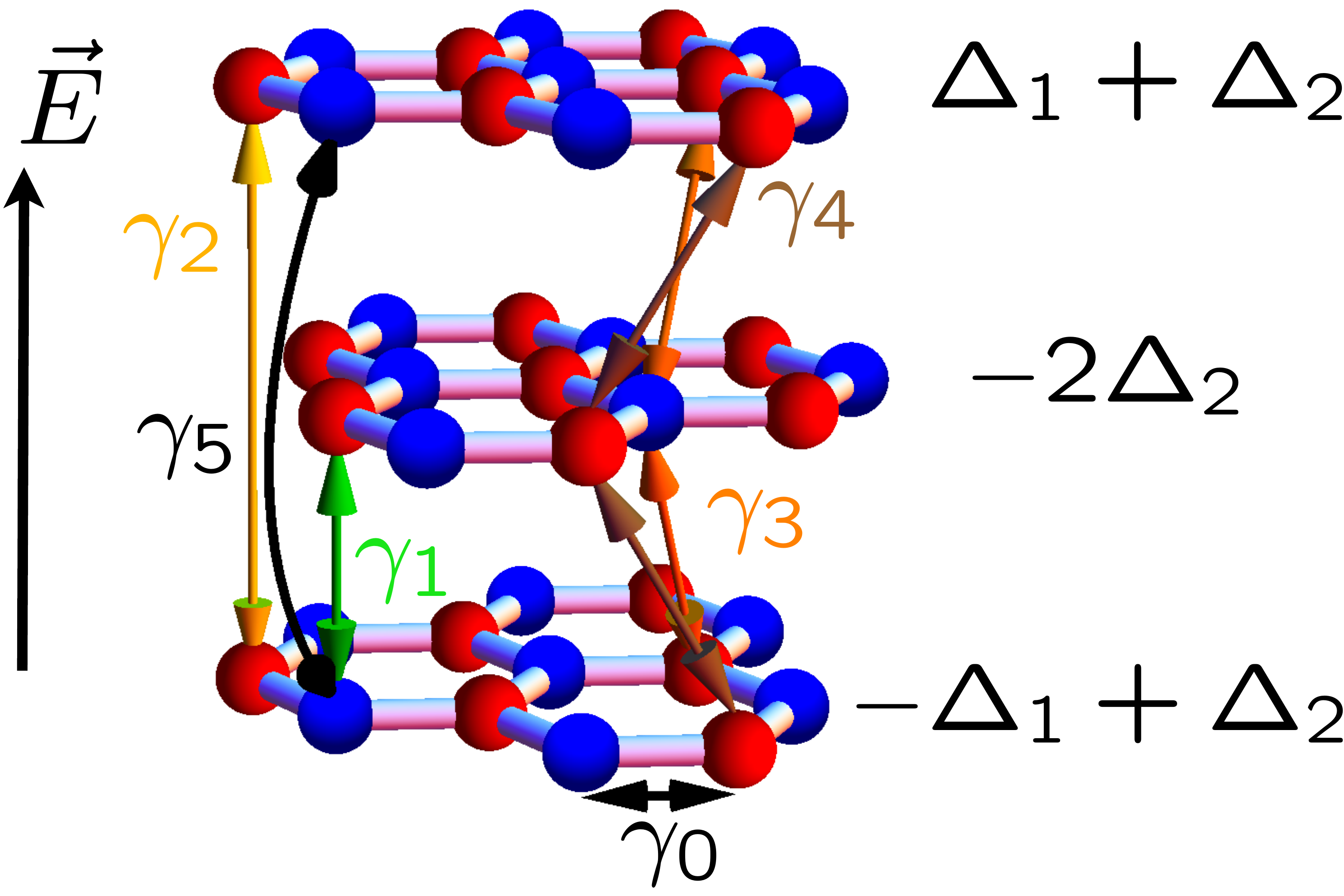}
\end{minipage}%
\hfill
\begin{minipage}[t]{0.25\columnwidth}
\mbox{}\par
\begin{tabular}{c r}
\hline\hline
par. & value\\
\hline
$\gamma_0$  & 3.1\,{\rm eV} \\ 
$\gamma_1$ &  0.39\, {\rm eV} \\
$\gamma_2$ &  -0.028\, {\rm eV} \\
$\gamma_3$ & 0.315\, {\rm eV} \\
$\gamma_4$ &  0.041\, {\rm eV} \\
$\gamma_5$ & 0.05\, {\rm eV} \\
$\delta$ & 0.046\, {\rm eV} \\
\hline\hline
\end{tabular}
\end{minipage}%
\hspace*{0.02\linewidth}
\caption{ \label{Fig:aba-lattice} (Color online)
Schematic view of the lattice structure of the $ABA$-stacked trilayer graphene (left) and values of corresponding tight-binding parameters adopted from Ref.~[\onlinecite{ThitiExp}] (right). Red (blue) atoms belong to $A$ ($B$) sublattice of the corresponding layer. }
\end{center}
\end{figure}

\subsection{Tight-binding model \label{Sec:TB}}
In order to describe the band structure of the unbiased trilayer graphene, we adopt the standard Slonczewski-Weiss-McClure parametrization of the tight-binding model.~\cite{Dress-tb} Corresponding tight-binding parameters are schematically illustrated in Fig.~\ref{Fig:aba-lattice}.  Six tight-binding parameters $\gamma_0\ldots\gamma_5$ describe hopping matrix elements between different atoms:
\begin{subequations} \label{Eq:gammas-def}
\bea
  A_i\leftrightarrow B_i : \gamma_0,\quad &\qquad&
  B_{1,3}\leftrightarrow A_{2} : \gamma_1,
  \\
  A_1\leftrightarrow A_3 : \frac12 \gamma_2,
  &\qquad&
  A_{1,3}\leftrightarrow B_2 :  \gamma_3,
  \\
\begin{matrix}
  A_{1,3}\leftrightarrow A_2\\
  B_{1,3}\leftrightarrow B_2
\end{matrix}:  -\gamma_4,
  &\qquad&
  \ \  B_1\leftrightarrow B_3 :  \frac12 \gamma_5,
\eea
\end{subequations}
where $A_i$ ($B_i$) refers to an atom from $A$ ($B$) sublattice, and index $i=1\ldots 3$ labels three layers~[see Fig.~\ref{Fig:aba-lattice} for notation of sublattices]. In addition, parameter $\delta$ accounts for an extra on-site potential for $B_1$, $A_2$, and $B_3$ sites,  which are on top of each other. 
The values of different tight-binding parameters are far from being settled. In particular, authors of Ref.~\onlinecite{ThitiExp} used experimentally determined  higher LLs crossings to refine values of tight binding parameters. Corresponding values are listed in Fig.~\ref{Fig:aba-lattice} and will be used in what follows.

Effect of an external perpendicular electric field (displacement field) can be described by including additional on-site potentials for different layers $U_{1}\ldots U_{3}$.~\cite{Lu-mult,Guinea-mult,MinAbInitio,McCannGateABA}. It is convenient to introduce parameters 
\begin{equation}   \label{Eq:D1D2def}
  \D_1  =  (-e)\frac{U_1-U_2}{2},
  \qquad
  \D_2
  =
  (-e)\frac{U_1-2U_2+U_3}{6}.
\end{equation}
Physically, parameter~$\D_1$ is responsible for the potential difference between top and bottom layers which is caused by the application of the displacement field to the sample. Parameter~$\D_2$ describes the deviation of potential on the middle layer from the mean of the potentials on top and bottom layers. As we shall discuss below, non-zero parameter~$\D_2$ can emerge when screening is taken into account and the potential distribution is calculated self-consistently. In principle, even \emph{without bias}, non-zero value of~$\D_2$ is allowed by symmetry. In what follows we will demonstrate that LL crossings pattern is very sensitive to precise value of~$\D_2$.

For our analysis below, it will be convenient to separate the Hamiltonian of the unbiased trilayer, $H_\text{0}$, from terms related to the potential imbalance between layers,
\begin{equation} \label{Eq:Htot}
  H
  =
  H_\text{0} + H_{\Delta_1}+H_{\Delta_2}. 
\end{equation}
In the momentum representation, the tight-binding Hamiltonian can be written in a compact matrix form, with different components corresponding to the wave function amplitudes on the six sublattices.  Let us choose the basis $\{A_1, B_1, A_2, B_2, A_3, B_3\}$. In this basis, the three terms in the Hamiltonian can be written as
\begin{equation}\label{Eq:Hhop}
  H_\text{0}
  =
  \bpm
  0 & \g_0 t_{\bm k}^*  & \g_4 t_{\bm k}^*   &\g_3 t_{\bm k} &\gamma_2/2 & 0\\
    \g_0 t_{\bm k} & \delta    &\g_1 &\g_4 t_{\bm k}^* &0          & \gamma_5/2\\
  \g_4 t_{\bm k}  & \gamma_1       &  \delta     &\g_0 t_{\bm k}^* & \g_4 t_{\bm k}    &\gamma_1\\
  \g_3 t_{\bm k}^* &  \g_4 t_{\bm k}    &  \g_0 t_{\bm k}      &0      &\g_3 t_{\bm k}^*    &\g_4 t_{\bm k}\\
  \gamma_2/2   &  0      & \g_4 t_{\bm k}^*   & \g_3 t_{\bm k}     &0          &\g_0 t_{\bm k}^* \\
  0  &   \g_5/2     & \g_1     &  \g_4 t_{\bm k}^*     &      \g_0 t_{\bm k}     & \delta \\
  \epm ,
\end{equation}
\begin{equation} \label{Eq:Hdiag}
 H_{\Delta_1}= {\rm diag}(\D_1, \D_1, 0, 0, -\D_1,-\D_1),
\end{equation}
\begin{equation} \label{Eq:Hdiag2}
 H_{\Delta_2}= {\rm diag}(\D_2, \D_2, -2\D_2, -2\D_2, \D_2,\D_2).
\end{equation}
In Eq.~(\ref{Eq:Hhop}) we used  $t_{\bm k}$ as a short-hand notation for a sum over nearest neighbors in a single honeycomb lattice: 
\begin{equation} \label{Eq:PI-def}
  t_{\bm k} = \sum_{j=1}^3 e^{i {\bm k}\cdot{\bm a}_j}
  =-1-2e^{\sqrt{3}i k_y/2}\cos\frac{k_x}{2},
\end{equation}
where ${\bm a}_1=(0,1/\sqrt3)$, ${\bm a}_{2,3}=(\mp 1/2,-1/2\sqrt{3})$, are vectors connecting an $A_1$ site to its nearest neighbors, and momenta are measured in units of inverse lattice constant, $a=2.46~\AA$. 

To obtain an effective low-energy Hamiltonian, we expand Eq.~(\ref{Eq:Hhop}) in the vicinity of $K_+$ and $K_-$ points, located at position $(\pm 4\pi/3, 0)$ in the hexagonal Brillouin zone. This expansion reduces to replacing $\gamma_i t_{\bm k} \to v_i \pi$, where
\begin{equation} \label{Eq:Pi-exp}
\Pn = \xi k_x+ik_y, \,\,  \hbar v_i= \frac{\sqrt{3}}{2}a \gamma_i,
\end{equation}
with $\xi=\pm1$ for $K_+$ and $K_-$ points, respectively.

\subsection{Effective Hamiltonian of unbiased trilayer graphene}

In the absence of external bias, the band structure of the Hamiltonian~(\ref{Eq:Htot}) consists of monolayer-like and bilayer-like bands, as shown by Koshino and McCann.~\cite{McCannParityValleyMultilayer} To illustrate the decoupling of the Hamiltonian, we consider a different basis, 
\be \label{e:basis2}
 \left\{ \frac{A_1-A_3}{\sqrt2}, \frac{B_1-B_3}{\sqrt2}, \frac{A_1+A_3}{\sqrt2}, B_2, A_2, \frac{B_1+B_3}{\sqrt2} \right\}.
\ee
In the new basis, sum of two terms, $H_{0}+H_{\Delta_2}$ ($\Delta_2$ is allowed by symmetry even in the absence of electric field), acquires a block-diagonal structure, 
\be
  H_0+H_{\Delta_2}  \label{Eq:Htotrot}
  =
  \bpm
   H_\text{SLG} & 0 \\ 
   0   &  H_\text{BLG}\\
  \epm,
\ee
where the monolayer-like and bilayer-like blocks are defined as
\begin{subequations}
\begin{gather} \label{Eq:HSLG}
  H_\text{SLG}
  =
\bpm
 \Delta_2-\frac{\gamma_2}{2} & v_0\Pc \\
 v_0 \Pn  & -\frac{\gamma_5}{2}+\delta +\D_2 \\
\epm,
\\ \label{Eq:HBLG}
  H_\text{BLG}
 =
\bpm
 \frac{\gamma_2}{2}+\D_2 & \sqrt{2} v_3 \Pn  & -\sqrt{2}v_4 \Pc  &
   v_0 \Pc  \\
  \sqrt{2} v_3 \Pc & -2 \D_2 & v_0 \Pn & -\sqrt{2}v_4 \Pn \\
  -\sqrt{2}v_4 \Pn  & v_0\Pc  & \delta -2 \D_2 & \sqrt{2} \gamma_1 \\
  v_0\Pn & -\sqrt{2}v_4 \Pc  & \sqrt{2} \gamma_1 & \frac{\gamma_5}{2}+\delta
   +\D_2
\epm.
\end{gather}
\end{subequations}
The effective Hamiltonian of bilayer-like sector can be simplified further by noting that the low-energy states predominantly reside on $A_1+A_3$ and on $B_2$ sublattices. The higher energy bands, which reside mostly on sublattices $B_1+B_3, A_2$ have a large band gap of order $\pm\sqrt{2}\gamma_1\sim0.5$~eV. Taking into account virtual excitations to these bands, one obtains an effective $2\times 2$ Hamiltonian of the bilayer-like band:~\cite{McCannFalkoLLBLG} 
\be
  H_\text{BLG}\approx
  H^{(0)}_\text{BLG}+H^{(1)}_\text{BLG},
\ee
where $H^{(0)}_\text{BLG}$ is the leading approximation,
\be \label{Eq:HBLG11}
  H^{(0)}_\text{BLG}
  =
  -\frac{1}{2m}\bpm 0 & {\Pc}^2\\ \Pn^2& 0\epm,
  \quad
  \frac{1}{2m} = \frac{v^2}{\sqrt{2}\gamma_1},
\ee
and  $H^{(1)}_\text{BLG}$ describes various corrections:
\begin{multline}\label{Eq:H11pert}
  H^{(1)}_\text{BLG}
  =
    \sqrt{2}v_3 \bpm 0 & \Pn\\ \Pc& 0\epm
    +
    \bpm \gamma_2/2+\Delta_2 & 0\\ 0& -2\Delta_2\epm
     \\
    +
    \frac{v^2}{2\gamma_1} \bpm  (\delta-2\Delta_2)\Pc\Pn & 0\\ 0& (\gamma_5/2+\delta+\Delta_2)\Pn\Pc\epm.
\end{multline}
Note that in the above expression for $H^{(1)}_\text{BLG}$ we omitted from terms which renormalize mass in $H^{(0)}_\text{BLG}$~[Eq.~(\ref{Eq:HBLG11})]. Also, the last term in Eq.~(\ref{Eq:H11pert}) does not include contributions which come with an additional small parameter $\gamma_1\gamma_4/(\gamma_0\delta)\approx 0.11$. 

\begin{figure*}[t]
\begin{center}
\setlength{\unitlength}{\columnwidth}
\begin{picture}(2,.85)
\put(-.05,0){
\includegraphics[width=1.22\columnwidth]{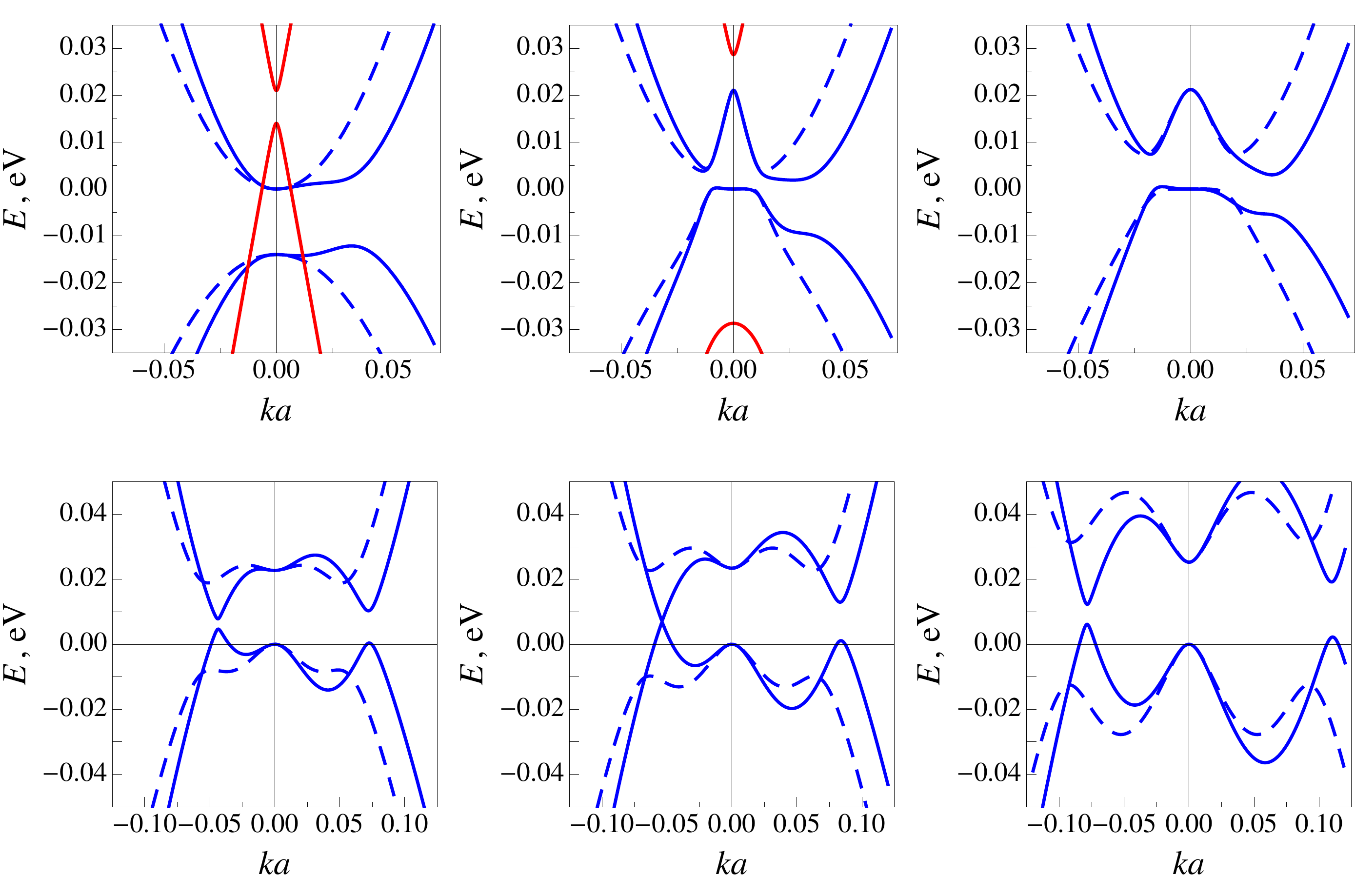}
\qquad 
\includegraphics[width=0.7\columnwidth]{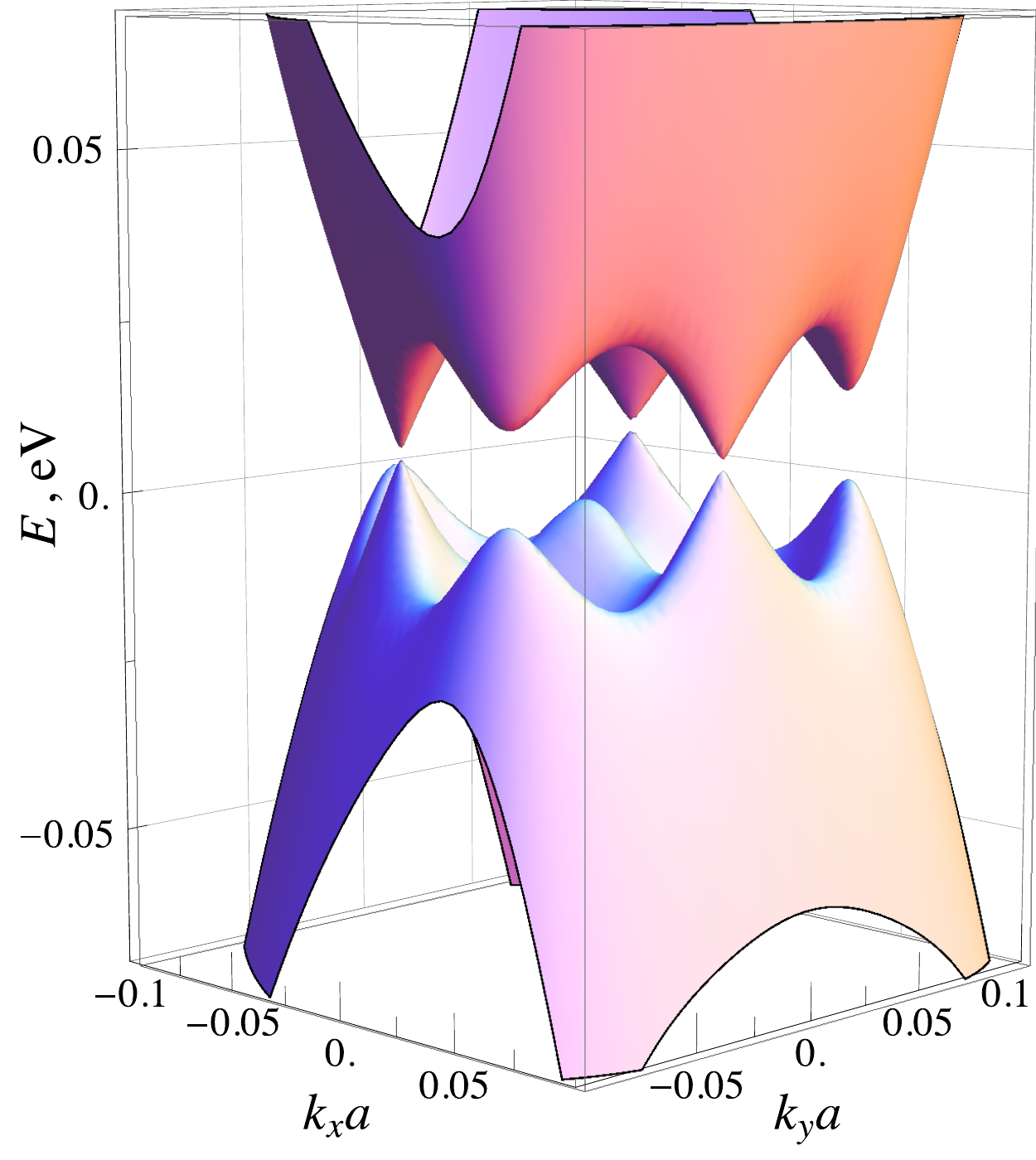}
}
\put(0.05, .81){(a)}
\put(0.15, 0.81){$\Delta_1 = 0$}
\put(0.46, .81){(b)}
\put(0.54, 0.81){$\Delta_1 = 0.025$~\text{eV}}
\put(0.85, .81){(c)}
\put(0.95, 0.81){$\Delta_2 = 0.05$~\text{eV}}
\put(0.05, .39){(d)}
\put(0.15, 0.39){$\Delta_1 = 0.15$~\text{eV}}
\put(0.46, .39){(e)}
\put(0.54, 0.39){$\Delta_1 = 0.18$~\text{eV}}
\put(0.85, .39){(f)}
\put(0.95, 0.39){$\Delta_2 = 0.25$~\text{eV}}
\put(1.32, .81){(g)}
\put(1.42, 0.81){$\Delta_1 = 0.2$~\text{eV}}
\end{picture}
\caption{ \label{Fig:ABAband} (Color online)
Evolution of the band structure of trilayer graphene with bias $\Delta_1$. The bands in the vicinity of $K_+$ point in the Brillouin zone are shown as a function of $ka$, where ${\bm k}$ describes momentum relative to the $K_+$ point. Solid (dashed) lines correspond to ${\bm k}$ parallel to $x$-axis ($y$-axis). At zero bias (panel (a)), red and blue lines describe monolayer-like and bilayer-like bands. At non-zero bias the monolayer-like and bilayer-like bands hybridize and no clear distinction can be made between them. As a result of this hybridization, one pair of bands moves to higher energies. The other pair of bands undergoes a series of transformations which lead to seven emergent Dirac points at higher bias $\Delta_1 \gtrsim 0.1\, {\rm meV}$. The positions of these Dirac points in the Brillouin zone are illustrated in panel (g). }
\end{center}
\end{figure*}
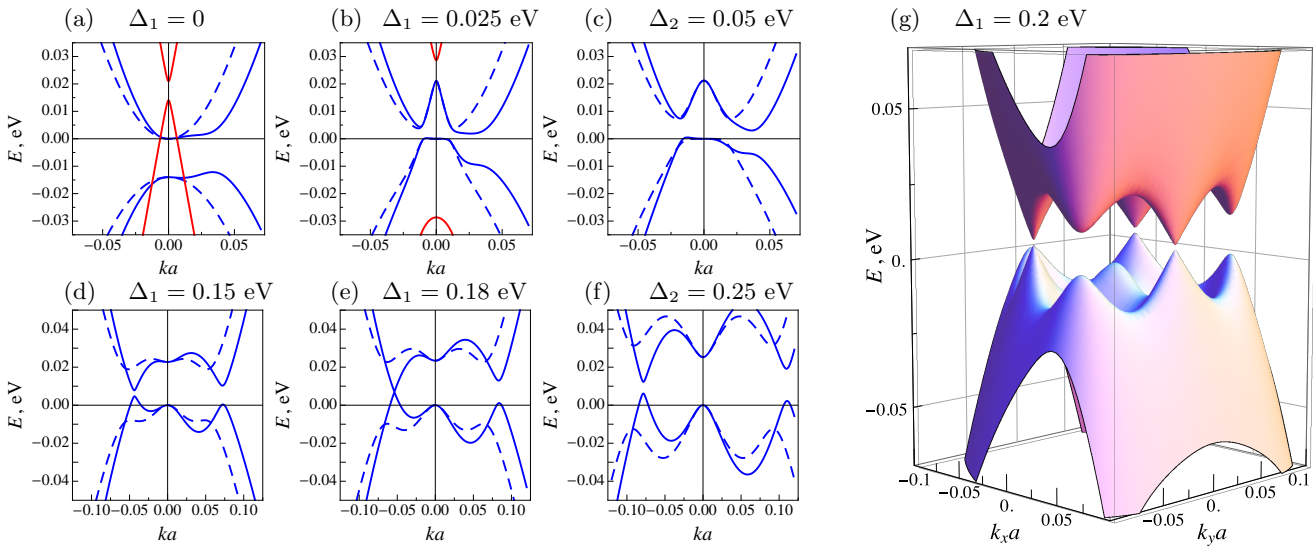

Therefore, in the limit when there is no electric field~($\D_1=0$), band structure of $ABA$ stacked trilayer graphene has SLG and BLG sectors which are decoupled from each other. The resulting band structure of unbiased trilayer is illustrated in Fig.~\ref{Fig:ABAband}(a).

When a displacement  field is applied, the reflection symmetry between the top and bottom layer is broken and the $H_{\Delta_1}$ appears in the Hamiltonian (note that electric field also affects $\Delta_2$, which, in general, should be calculated self-consistently, see below). In the rotated basis~(\ref{e:basis2}) this term corresponds to hybridization between the monolayer-like and bilayer-like blocks and has a form: 
\be
  H_{\D_1}= \bpm
 0 & H_{\rm hyb}  \\
 H_{\rm hyb}^T & 0  \\
\epm,
\quad
H_{\rm hyb}
  =
\bpm
 \D_1 & 0 & 0 & 0 \\
 0 & 0 & 0 & \D_1 \\
\epm.
\ee
As a next step, we will discuss the evolution of the bands resulting from this hybridization. 

\section{Band structure evolution under bias: a new set of Dirac points  \label{Sec:Band}}

An electric field completely changes the low-energy properties of trilayer graphene, hybridizing the monolayer and bilayer-like bands. In this Section we shall study how the bands transform as bias increases. We will see that at relatively small bias, there is a complex interplay of terms induced by electric field and smaller trigonal warping terms in the Hamiltonian, which gives rise to rather flat low-energy bands. At higher bias ($\Delta_1 \gtrsim 0.1 {\rm eV}$), a new set of Dirac points emerges: there is a total of seven DPs in each valley, one of them situated at $K_{\pm}$, and two groups of three related by rotational symmetry. The positions, mass gaps, as well as velocities of these DPs change as a function of bias.  

Since analytical treatment is no longer possible, we resort to numerical diagonalization of the tight-binding Hamiltonian (\ref{Eq:Hhop}). The evolution of the band structure in valley $K_+$ as $\D_1$ is changed from zero to $0.25$~eV is illustrated in Fig.~\ref{Fig:ABAband}. Notice that the band structure in $K_-$ valley is related to that in $K_+$ valley by the time-reversal symmetry. Throughout this Section, we will put $\Delta_2=0$, because $\Delta_2$ is expected to be of the order of several ${\rm meV}$, and we found that it cannot qualitatively change the band structure. We emphasize, however, that $\Delta_2$ will play an important role for LL spectrum and LL crossing pattern in Section~\ref{Sec:LLs}. 

Band structure in the absence of bias, Fig.~\ref{Fig:ABAband}(a), consists of a monolayer-like band (shown in red) and a trigonally warped bilayer-like band (shown in blue). Relatively small bias $\Delta_1\approx 0.025 \, {\rm eV}$ in Fig.~\ref{Fig:ABAband}(b), turns out to be sufficient to hybridize bands; an electron-like and a hole-like bands which used to be of monolayer type, drift away to higher energies. The band gap for these bands becomes of the order $0.03 \, {\rm eV}$ already at $\Delta_1=0.025\, {\rm eV}$, which means that they will be essentially irrelevant for the low-energy properties. 

As the bias is increased further ($\Delta_1=0.05\, {\rm eV}$ in Fig.~\ref{Fig:ABAband}(c)), the two remaining bands stay close to the neutrality point, and they are separated by a small band gap of the order of a few tens of ${\rm meV}$. Near the $K$-points, the bands have a complicated dispersion relation and trigonal warping is important. One notable feature is that the hole-like band is nearly flat in the vicinity of the $K_+$ point. 

When $\Delta_1$ becomes larger than $0.1\, {\rm eV}$, the evolution of the bands leads to seven new Dirac points: one at the $K_+$ point, and six situated off-center. The off-center DPs consist of two groups of three, related by three-fold rotational symmetry. All Dirac points are generally split by a small mass. The locations and splittings of the emergent DPs for $\Delta_1=0.2\, {\rm eV}$ are illustrated in Fig.~\ref{Fig:ABAband}(g). In the interval of values $\Delta_1=0.15\div0.25\, {\rm eV}$, the off-center DPs move further away from the $K_+$ point. Also, for one of the groups the band gap closes and the Dirac mass changes sign at $\Delta_1\approx 0.18\, {\rm eV}$ [see Fig.~\ref{Fig:ABAband}~(d)-(g)]. The velocity also changes, although not very significantly. 

To understand the origin of the new DPs and some of their properties, including their position in the Brillouin zone, let us re-inspect the tight binding model at large bias. We note that at large bias, the most important terms in the band structure are $\gamma_0, \gamma_1, \gamma_3$ and $\Delta_1$. It is convenient to separate the Hamiltonian into the leading part which contains only these terms, $H^{(0)}$, and the remaining subleading part,  $H^{(1)}$: 
\be\label{eq:H_separate}
H= H^{(0)}+ H^{(1)}.  
\ee
To elucidate the structure of  $H^{(0)}$ let us re-arrange the basis (\ref{e:basis2}) as follows: 
\be \label{e:basis3}
 \left\{ \frac{A_1-A_3}{\sqrt2}, B_2, \frac{B_1+B_3}{\sqrt2}, \frac{A_1+A_3}{\sqrt2}, A_2, \frac{B_1-B_3}{\sqrt2} \right\}.
\ee
Using Eqs.~(\ref{Eq:Hhop})-(\ref{Eq:Hdiag}), we write $H^{(0)}$ in the vicinity of $K_+$ point  in the new basis: 
\be\label{eq:Hlead}
  H^{(0)}=
  \bpm
   0 & C \\ 
   C^{\dagger}  & 0 \\
  \epm, 
  \quad 
  C=
\bpm
   \Delta_1 & 0  & v_0 \pi^\dagger \\ 
  \sqrt{2} v_3 \pi^\dagger   &  v_0 \pi  & 0 \\
    v_0\pi & \sqrt{2} \gamma_1 & \Delta_1 \\
  \epm. 
\ee
Similarly, one can find an explicit form of  subleading part of $H^{(1)}$, but we will not need it here.

The chiral symmetry of the Hamiltonian~(\ref{eq:Hlead}) greatly simplifies the analysis of its band structure. This symmetry implies that the spectrum is particle-hole-symmetric. We find that for any $\Delta_1$, $H^{(0)}$ has seven massless DPs. Their positions can be found from condition $\mathop{\rm det} C=0$, which reduces to an algebraic equation on $\pi$: 
\be\label{eq:pi_equation}
\Delta_1^2 v_0 \pi+v_0 \pi^* (2 \gamma_1 v_3 \pi^*-v_0^2 \pi^2)=0. 
\ee
First of all, $\pi=0$ is an obvious root of the above equation. Second, introducing parametrization $\pi=p e^{i\theta}$, we obtain the other roots which give the positions of the remaining six off-center DPs: 
\begin{eqnarray}\label{eq:roots1}
p_1
&=&
\frac{\sqrt{\gamma_1^2 v_3^2+\Delta_1^2v_0^2}+\gamma_1 v_3}{v_0^2}, 
\quad \theta=0, 2\pi/3, 4\pi/3, 
\\ \label{eq:roots2}
p_2
&=&
\frac{\sqrt{\gamma_1^2 v_3^2+\Delta_1^2v_0^2}-\gamma_1 v_3}{v_0^2}, 
\quad \theta=\pi/3, \pi, 5\pi/3. 
\end{eqnarray}
Therefore, at small $\Delta_1 \ll \gamma_1 v_3/v_0 \approx 0.04\, {\rm eV}$ the off-center DPs remain relatively close to the $K_+$ point and their position depends approximately quadratically on~$\Delta_1$: $p_1\approx {2\gamma_1 v_3 }/{v_0^2} +{\Delta_1^2}/({2 \gamma_1 v_3}) $,  $p_2 \approx {\Delta_1^2}/({2 \gamma_1 v_3})$. At large bias, $\Delta_1 \gg 0.04\, {\rm eV}$, the distance of the off-center DPs from $K_+$ grows linearly with $\Delta_1$, $p_{1(2)}\approx ({\Delta_1 v_0 \pm \gamma_1 v_3})/{v_0^2}$. 

\begin{figure}
\begin{center}
\includegraphics[width=1\columnwidth]{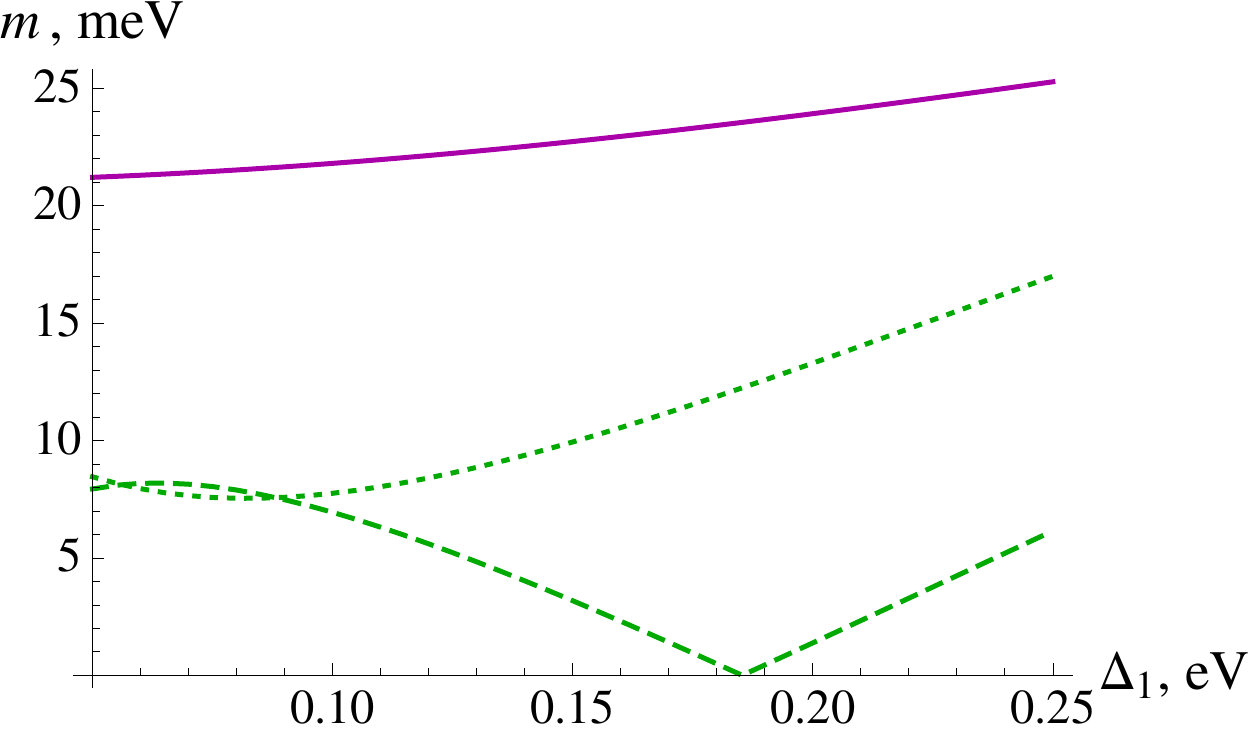}\\
\caption{\label{fig:mass} (Color online)
Evolution of the mass of emergent Dirac fermions as a function of bias for central DP (purple), first group of off-center DPs (dotted green) and second group of off-center DPs (dashed green). The mass of the second group of DPs vanishes at $\Delta_1\approx 0.185 \, {\rm eV}$ and becomes negative at a larger bias.}
\end{center}
\end{figure}

The chiral symmetry is broken by the subleading part of the Hamiltonian, $ H^{(1)}$ in Eq.~(\ref{eq:H_separate}). Consequently, all seven species of emergent Dirac fermions become massive, as is evident from Fig.~\ref{Fig:ABAband}. We extract the Dirac mass as a function of bias for these DPs from numerically evaluated spectrum.~\footnote{It is also possible to obtain the velocity and mass of emergent Dirac fermions analytically.~\cite{Koshino12}} The result is illustrated in Fig.~\ref{fig:mass}. The most interesting feature is that the mass of the second group of Dirac fermions vanishes at $\Delta^*_1\approx 0.185\, {\rm eV}$, and changes sign for larger values of $\D_1$. Note, that the value of $\Delta_1^*$ depends on the tight-binding parameters. The mass of the first group of Dirac fermions grows monotonically, changing from $8\, {\rm meV}$ at $\Delta_1=0.1\, {\rm eV}$ to $18\, {\rm meV}$ at $\Delta_1=0.25\, {\rm eV}$. Finally, the mass of the central DP depends on bias relatively weakly. 

The tunability of the Dirac masses will manifest itself in a characteristic non-monotonic bias dependence of conductivity at the neutrality point. At $\Delta_1< \Delta_1^*$ and $\Delta_1> \Delta_1^*$ we expect an activated temperature dependence of the conductivity with an activation gap set by the smaller of the Dirac masses. When $\Delta_1 \approx \Delta_1^*$, assuming the disorder is weak, the conductivity will have a metallic temperature dependence and will be dominated by the twelve species of massless Dirac fermions (including spin and valley  degeneracy), becoming of the order of $12\times e^2/h$ (we assumed that a single Dirac specie has a conductivity approximately equal to $e^2/h$~\cite{CastroNeto09}).

We note that, as previously pointed out by McCann and Koshino,~\cite{McCannGateABA,KoshinoScreeningABA} the screening can be significant, and therefore the parameters $\Delta_1, \Delta_2$ should be calculated self-consistently. We have calculated the self-consistent values of $\Delta_1$, $\Delta_2$ as a function of external bias in the Hartree approximation, obtaining results which were consistent with Refs.~[\onlinecite{McCannGateABA},~\onlinecite{KoshinoScreeningABA}]. Interestingly, there is non-zero $\Delta_2\approx 2\, {\rm meV}$ even at zero external bias. This term will be important in our analysis of low-lying LLs in the following Section. 

\section{Landau levels and their crossings \label{Sec:LLs}}

In this Section, we analyze Landau level spectrum of the biased trilayer graphene. As we shall see, the bias-induced band structure transformations lead to a rich pattern of LL crossings. At high bias, when the new Dirac points are formed, corresponding three-fold degenerate groups of LLs emerge. This gives rise to an anomalous sequence of quantum Hall plateaus. At smaller bias, all LL degeneracies are lifted,  and therefore all quantum Hall plateaus are spaced by $e^2/h$. An interesting feature of the LL crossings pattern is its extreme sensitivity to certain tight-binding parameters, including $\Delta_2$ and $\gamma_3$. We explore how the existence and positions of various crossings depend on the values of these parameters, and argue that in an experiment the LLs crossing pattern can be used to determine the precise values of tight-binding parameters. 

\subsection{Zero Landau levels without displacement field \label{landau_no_bias}}
\begin{figure*}
\begin{center}
\setlength{\unitlength}{\columnwidth}
\begin{picture}(2,.9)
\put(-.05,0)
{
\includegraphics[width=0.645\columnwidth]{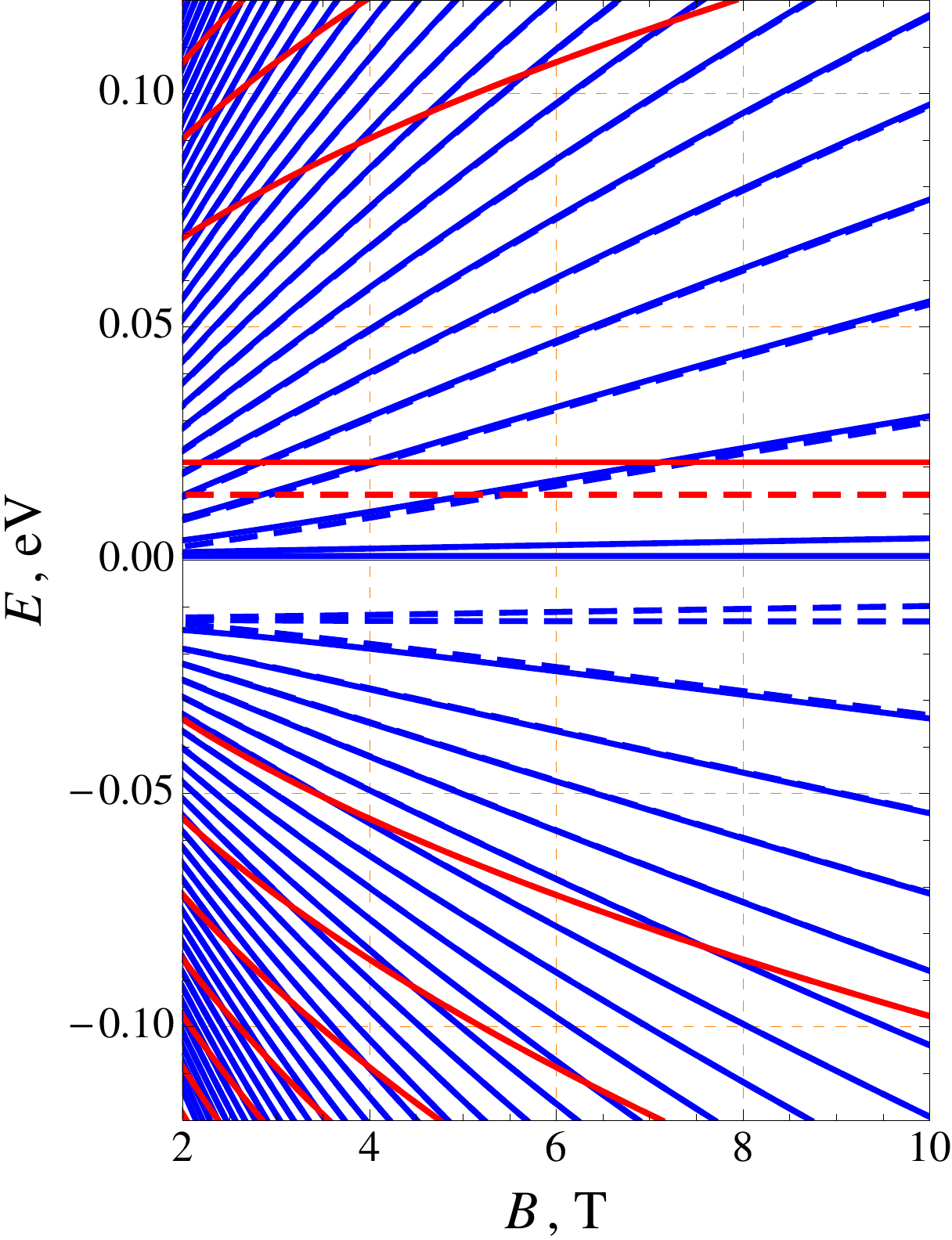}
\quad
\includegraphics[width=0.66\columnwidth]{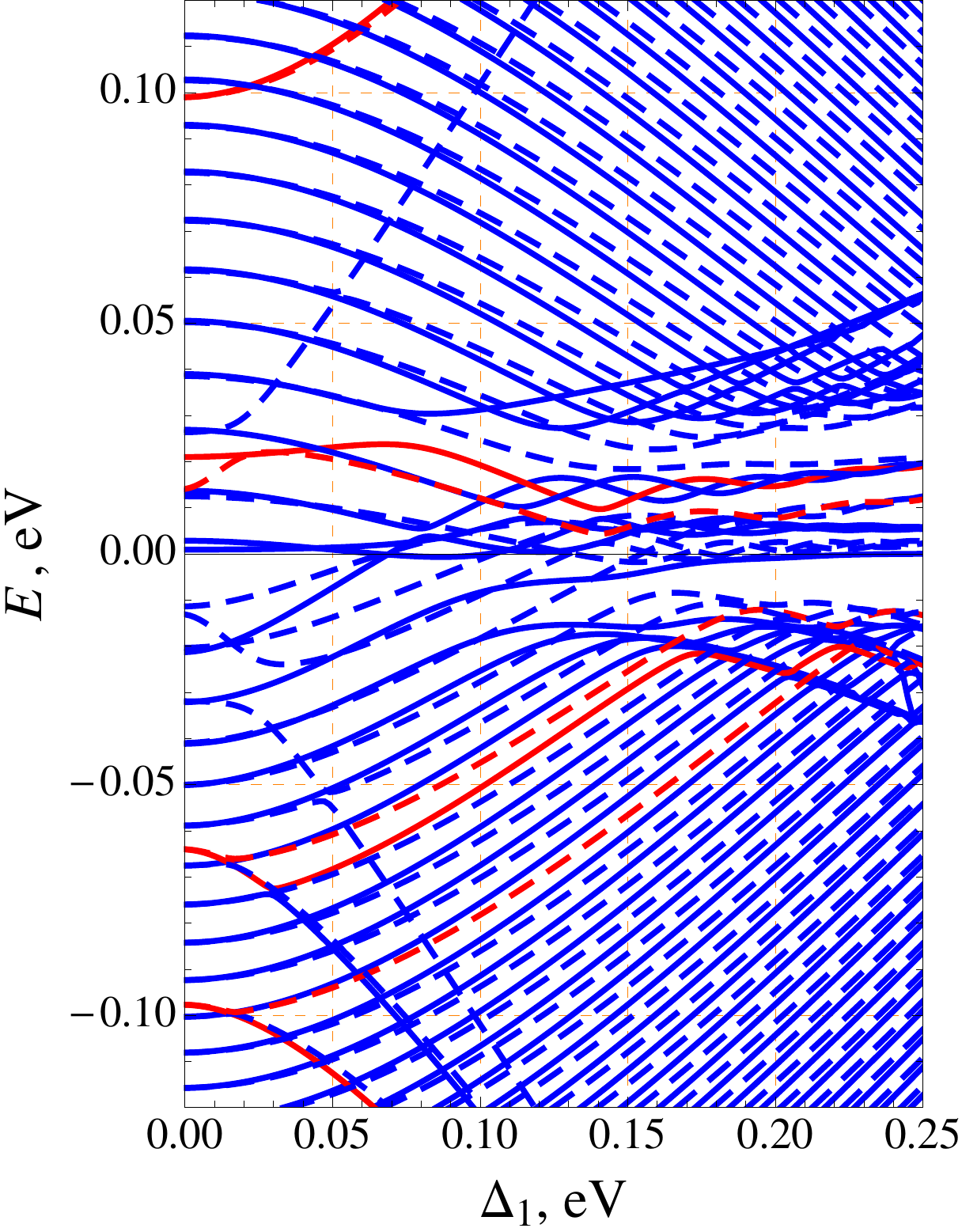}
\quad
\includegraphics[width=0.66\columnwidth]{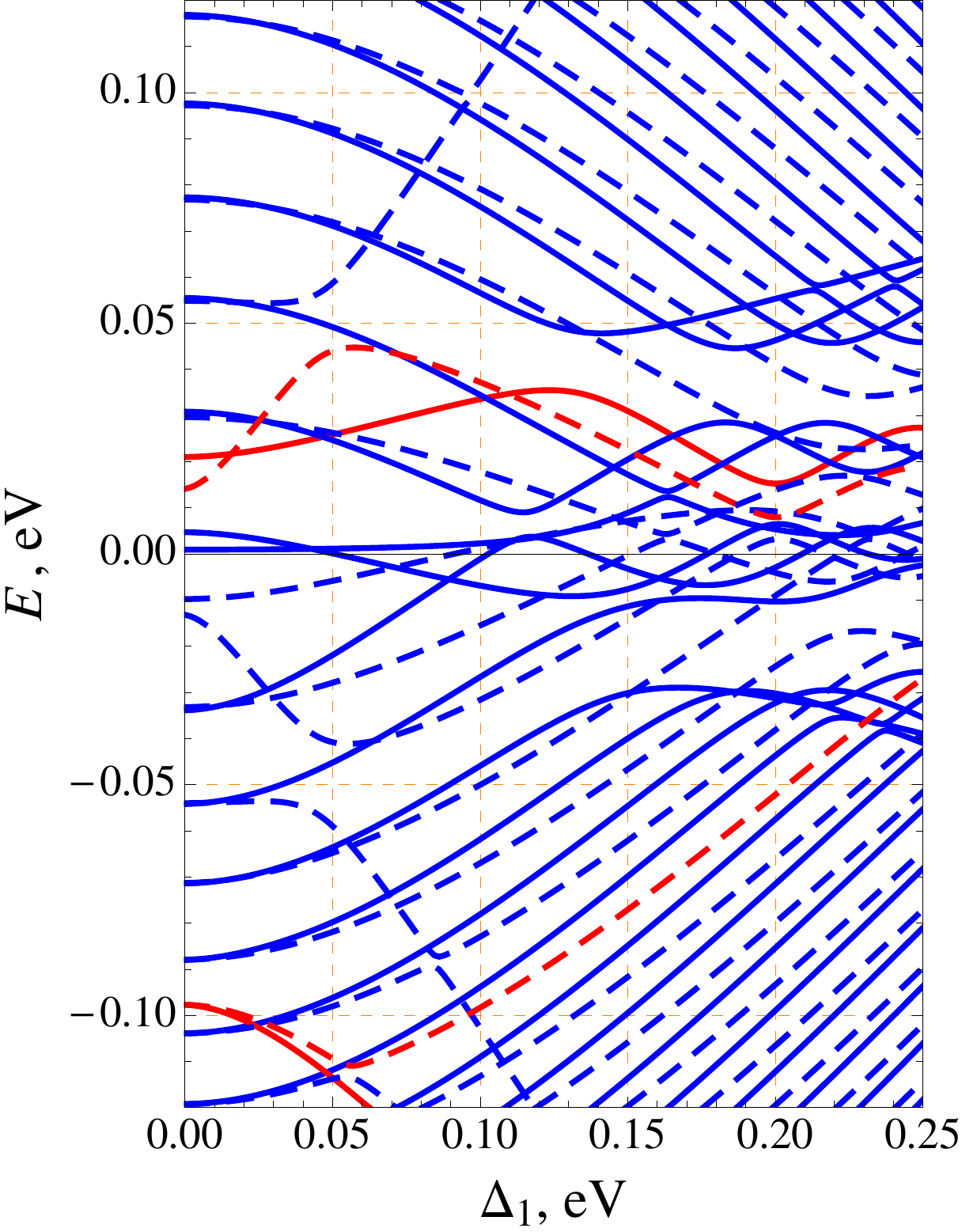}
}
\put(0.05, .86){(a)}
\put(0.28, 0.86){$\Delta_1 = 0$}
\put(0.75, 0.86){(b)}
\put(0.95, 0.86){$B = 5$~\text{T}}
\put(1.45, 0.86){(c)}
\put(1.65, 0.86){$B = 10$~\text{T}}
\put(0.6, 0.82){\blue {\footnotesize $6_\pm$}}
\put(0.6, 0.76){\blue {\footnotesize $5_\pm$}}
\put(0.6, 0.7){\blue {\footnotesize $4_\pm$}}
\put(0.6, 0.64){\blue {\footnotesize $3_\pm$}}
\put(0.6, 0.56){\blue {\footnotesize $2_\pm$}}
\put(0.6, 0.525){\red {\footnotesize $0_+$}}
\put(0.6, 0.495){\red {\footnotesize $0_-$}}
\put(0.6, 0.47){\blue {\footnotesize $1_+$}}
\put(0.6, 0.445){\blue {\footnotesize $0_+$}}
\put(0.6, 0.42){\blue {\footnotesize $1_-$}}
\put(0.6, 0.395){\blue {\footnotesize $0_-$}}
\put(0.6, 0.35){\blue {\footnotesize $2_\pm$}}
\put(0.6, 0.28){\blue {\footnotesize $3_\pm$}}
\put(0.6, 0.23){\blue {\footnotesize $4_\pm$}}
\put(0.6, 0.18){\blue {\footnotesize $5_\pm$}}
\put(0.6, 0.15){\red {\footnotesize $1_\pm$}}
\put(0.6, 0.115){\blue {\footnotesize $6_\pm$}}
\put(0.6, 0.07){\blue {\footnotesize $7_\pm$}}
\end{picture}
\\
\caption{ 
\label{Fig:LLnumer} (Color online)
Landau level spectrum as a function of magnetic field and bias. Plot (a) shows LL spectrum as a function of magnetic field $B=2\ldots 10$~Tesla at zero displacement field, $\D_1=0$, and $\D_2=0$. Landau levels from the monolayer-like and bilayer-like block are shown in red and blue color; solid (dashed) lines correspond to $K_+$ ($K_-$) valleys. LLs are labelled with their indices, e.g. ${\blue 1_+}, \red 1_+$ ($1^\text{b}_+, 1^\text{m}_+$ in the main text) stands for LL with index 1 from $K+$ valley of bilayer and monolayer sector respectively. Lifting of spin degeneracy by the Zeeman field is not shown, so that every LL is doubly degenerate.  Neutrality point is located between ${\blue 0_+}$ and ${\blue 1_+}$ LLs. Plots (b)-(c) show Landau levels at $B=5$~T and $B=10$~T as a function of displacement field. Red (blue) color is used to indicate LLs which belong to monolayer (bilayer) sector at $\Delta_1=0$.
}
\end{center}
\end{figure*}
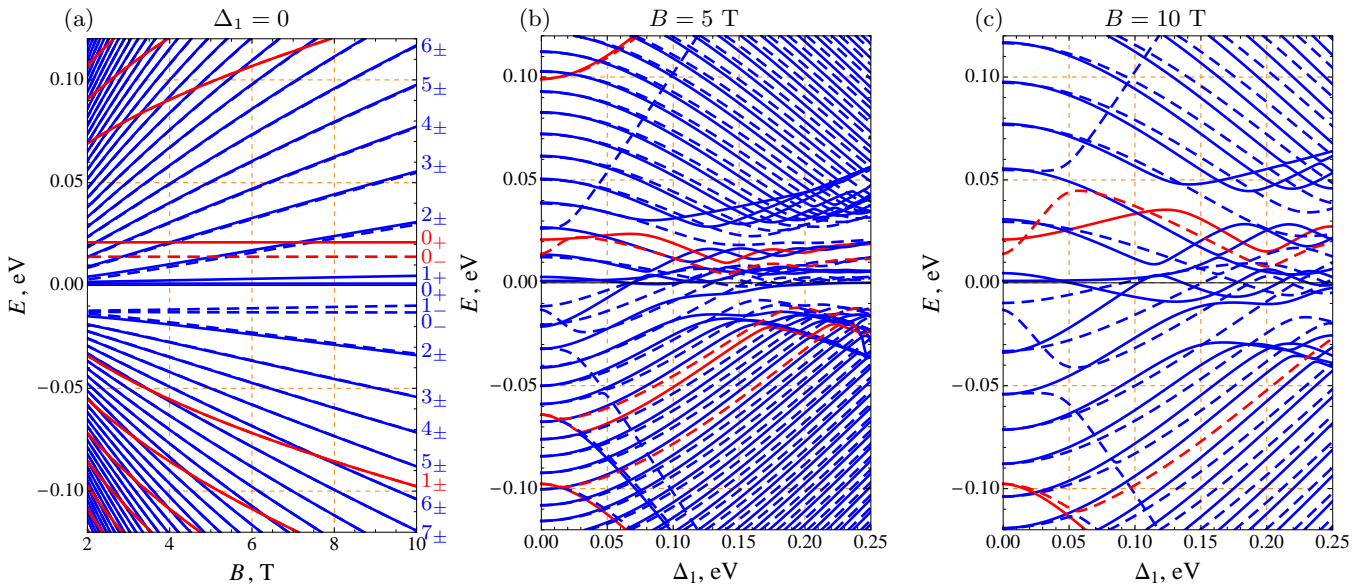

We start with reviewing known results about LL spectrum in the absence of bias.~\cite{McCann-LL,McCannParityValleyMultilayer,Katsnelson-LL,Peeters-LL} In what follows, we shall be mostly interested in the low-lying LLs. The energies of such LLs in the unbiased case can be obtained analytically~\cite{McCann-LL,Katsnelson-LL} from the effective low-energy model described in Section~\ref{Sec:TB}.  This will facilitate the identification of various low-lying LLs in numerical simulations, and also will help us to develop an intuition about their sensitivity to tight-binding parameters. 

In the absence of bias, the monolayer-like [Eq.~(\ref{Eq:HSLG})] and bilayer-like [Eqs.~(\ref{Eq:HBLG11})-(\ref{Eq:H11pert})] bands are independent. To find the spectrum of LLs, we adopt the Landau gauge, 
$A_x=0$, $A_y=Bx$, and make a substitution $\pi \to \pi-e(A_x+iA_y)$ in Eqs.~(\ref{Eq:HSLG}) and (\ref{Eq:HBLG11})-(\ref{Eq:H11pert}). Momentum $k_y$ is conserved (and is proportional to the guiding center position $X$), and $\pi$-operator acquires the following form ($\xi=\pm 1$ for $K_{\pm}$ valleys): 
\begin{equation} \label{Eq:pixX}
\pi=- \frac{i\hbar}{l_B} (\xi \partial_x+x-X),
\end{equation}
where the coordinate and momentum are measured in units of magnetic length $l_B=\sqrt{\hbar/eB}$ and $1/l_B$, respectively. We see that $\pi$ and $\pi^\dagger$ can be viewed as lowering and raising operators of the magnetic oscillator. In the basis of non-relativistic LL orbitals $|n\rangle$ (at a given $X$), the matrix elements of $\pi, \pi^\dagger$ in the two valleys are given by,  
\begin{subequations} \label{Eq:PnPcmatrix}
\begin{eqnarray}
  \nn
  K_+:\ &\Pn \LL{n} &= \frac{i\hbar}{l_B}\sqrt{2(n+1)}\LL{n+1},
  \\
  & \Pc\LL{n} &= -\frac{i\hbar}{l_B}\sqrt{2n}\LL{n-1},
  \\ \nn
  K_-:\ &\Pn\LL{n} &= \frac{i\hbar}{\l_B}\sqrt{2n}\LL{n-1},\\
  & \Pc\LL{n} &= -\frac{i\hbar}{l_B}\sqrt{2(n+1)}\LL{n+1}. 
\end{eqnarray}
\end{subequations}

The Hamiltonian of the monolayer-like band, Eq.~(\ref{Eq:HSLG}), describes two-dimensional Dirac-like fermions with a mass. Although the entire spectrum of Landau levels can be found analytically for the monolayer block, for our purposes it suffices to consider only the zeroth LLs in $K_{\pm}$ valleys:
\begin{subequations} \label{Eq:0LLSLG}
\begin{eqnarray} \label{Eq:0LLSLGK+}
 \Psi_{0^\text{m}_+}=
  \bpm 
  0\\
  |0\rangle
  \epm
  &,&
  \quad
    E_{0^\text{m}_+} =  \delta-\frac12\gamma_5+\Delta_2,
  \\  \label{Eq:0LLSLGK-} 
   \Psi_{0^\text{m}_-}=
  \bpm 
  |0\rangle \\
  0 
  \epm
  &,&
  \quad
 E_{0^\text{m}_-} = -\frac12\gamma_2+\Delta_2.
\end{eqnarray}
\end{subequations}
Therefore, zeroth LL in monolayer sector is valley split, with the energy difference given by $\D^\text{m}_\pm=E_{0^\text{m}_+}-E_{0^\text{m}_-}= \delta-\gamma_5/2+\gamma_2/2 \approx 7~\text{meV}$ (where we used tight-binding parameters listed in Fig.~\ref{Fig:aba-lattice}). This situation should be contrasted with monolayer and bilayer graphene, where LLs are valley-degenerate, at least at the single-particle level.~\cite{CastroNeto09}  

Finding low-lying LLs in the bilayer block is more challenging. In general, the trigonal warping described by the first term in~(\ref{Eq:H11pert}) does not allow to solve for LL spectrum analytically, as it breaks rotational symmetry and admixes arbitrarily high LLs. However, it turns out that trigonal warping gives relatively small corrections to LL energies, at least when magnetic field is not too small.~\cite{McCann-LL} Therefore, one can find LL spectrum in the bilayer block neglecting trigonal warping, which gives 
\begin{subequations} \label{Eq:0LLBLG}
\begin{eqnarray}  \label{Eq:0LLBLGK+}
 \Psi_{0^\text{b}_+} = 
   \bpm 
  0 \\
  |0\rangle 
  \epm
  &,& \ 
  E_{0^\text{b}_+} = -2\Delta_2,\
   \\ \label{Eq:1LLBLGK+}
    \Psi_{1^\text{b}_+} = 
     \bpm 
  0 \\
  |1\rangle 
  \epm
  &,& 
   \ 
   E_{1^\text{b}_+} = -2\Delta_2+\zeta \left( \frac{\gamma_5}{2}+\delta+\Delta_2\right)
 \\
 \Psi_{0^\text{b}_-} = 
     \bpm 
  |0\rangle \\
  0 
  \epm
  &,& \ 
   E_{0^\text{b}_-} = \frac{\gamma_2}{2}+\Delta_2,\
  \\  \label{Eq:0LLBLGK-}
    \Psi_{1^\text{b}_-} = 
        \bpm 
  |1\rangle \\
  0 
  \epm
  &,& \ 
    E_{1^\text{b}_-} = \frac{\gamma_2}{2}+\Delta_2+ \zeta (\delta-2\Delta_2),
\end{eqnarray}
\end{subequations}
where $\om_c={eB}/{m}={\sqrt2 e B v^2}/{\gamma_1}$ is the cyclotron frequency, 
and $\zeta= {\hbar\om_c}/({\sqrt2 \gamma_1})$ is a dimensionless parameter proportional to the magnetic field~($\zeta\approx 0.02 $ for $B=5\, {\rm Tesla}$) that controls splitting between two LLs in the same valley. The inter-valley splitting, approximately given by $\gamma_2/2+3\Delta_2$, is of the order of $10\, {\rm meV}$; it is very sensitive to the value of $\Delta_2$, since it enters with a factor of 3. Below we will see that even relatively small variations of $\Delta_2$ can completely change the pattern of LL crossings at small negative filling factors. Also, notice that the inter-valley splitting is much larger than the intra-valley splitting: the latter is proportional to $\zeta$, and is of the order $1\, {\rm meV}$ at $B=5\, {\rm T}$. 

\subsection{Landau levels at high bias}
We now proceed with the analysis of LL spectrum at a non-zero bias. As bias admixes the monolayer-like and bilayer-like blocks, and trigonal warping becomes important, approximate analytical treatment is no longer possible. Therefore, we find LL spectrum and wave functions numerically. 

If we fix guiding center position at some value, our Hamiltonian can be viewed as an infinite matrix. It is easily written in the basis of non-relativistic LL orbitals $|n\rangle$, using matrix elements of $\pi,\pi^\dagger$ operators from Eq.~(\ref{Eq:PnPcmatrix}). We truncate this infinite matrix, restricting the Hilbert space to LL orbitals with indices $n\leq \Lambda$. Finally, the problem of finding LL spectrum reduces to diagonalizing a matrix in which each element of $6\times 6 $ Hamiltonian, Eq.~(\ref{Eq:Htotrot}), is replaced by a matrix of dimensions $\Lambda\times \Lambda$.  Momentum operators are replaced by matrices corresponding to raising and lowering operators according to Eq.~(\ref{Eq:PnPcmatrix}) with the cutoff $\Lambda$. All other non-zero elements  $\gamma_i$ are replaced by $\gamma_i \mathds{1}_{\Lambda\times \Lambda}$. 

One subtlety associated with this procedure is due to the fact that the cutoff imposed on the ladder operators gives rise to new unphysical eigenvalues.~\footnote{Extra Landau levels are introduced by cutoff imposed on ladder operators. Indeed, it is impossible to realize the algebra of raising and lowering operators, $[a,a^\dagger]=1$ with matrices of finite size: for any finite matrices, $A$ and $B$ $\tr[A,B]=0$. This fact manifest itself in appearance of unphysical Landau levels which has to be removed.} 
Such unphysical eigenvalues has to be identified and removed. We do this by using the fact that, although these ``false'' Landau levels have low energy, their wave functions are dominated by Landau level orbitals with large indices (near cutoff $\Lambda$). We have found that cutoff $\Lambda\gtrsim30$ is sufficient to faithfully represent the evolution of LLs in vicinity of neutrality point with magnetic field and bias. All simulation presented in this paper are done with $\Lambda=50$. 

The evolution of LLs with magnetic field at zero bias is shown in Fig.~\ref{Fig:LLnumer}~(a). The zeroth LLs in bilayer and monolayer block can be easily identified in this picture because they disperse very weakly with magnetic field. These results are consistent with Ref.~[\onlinecite{McCann-LL}], however, notice, that we have chosen a different set of tight-binding parameters extracted from experiment~\cite{ThitiExp}, which leads to visible differences in the LL spectrum. 

Landau level spectrum as a function of bias for fixed values of magnetic field, $B=5\, {\rm T}$ and $B=10\, {\rm T}$,  illustrated in Fig.~\ref{Fig:LLnumer}(b)-(c), reveals a very complex pattern of crossings. At smaller field $B=5\, {\rm T}$ and large bias $\Delta \gtrsim 0.15~{\rm eV}$, both the dispersion and degeneracy of LLs changes: in particular, there are three-fold-degenerate groups which are associated with the new emergent Dirac points in the band structure. In addition, there is a singly-degenerate zeroth LL that originates from the central Dirac point in each valley. The dispersion of new zeroth LLs is determined by the dependence of the Dirac mass on the bias~(see Fig.~\ref{fig:mass}). 

It should be noted that the three-fold degeneracy of a subset of DPs gives rise to the three-fold degenerate groups of LLs only when magnetic field is not too strong. This means that the inverse magnetic length must be smaller than the distance between the DPs in the BZ. Considering $\Delta_1 \gg 0.04 \, {\rm eV}$ and using the approximate expressions for the DP positions obtained in the previous Section, we write this condition as follows, ${\hbar}/{l_B}\ll {\Delta_1}/{v_0}$. Thus, three-fold degenerate groups of LLs are most easily observable at smaller fields. 

Experimentally, the degeneracy of certain LLs at high bias (and relatively small field) will manifest itself in an anomalous quantum Hall sequence. Some of the steps will be characterized by a 
jump of Hall conductivity by $3e^2/h$ (when a three-fold-degenerate LL is filled), while others will correspond to a jump of $e^2/h$ (when a singly degenerate LL is filled). Here we assumed that the spin degeneracy is lifted by the Zeeman interaction. In experiment, such an anomalous quantum Hall sequence will signal the formation of new DPs.
 
 \subsection{Anti-crossings and the role of $\gamma_3$}

As we discussed above, bias gives rise to three-fold degenerate groups of LLs, which is in full agreement with the emergencece of low-energy DPs in the band structure. Another notable feature of a LLs pattern, which is visible in Fig.~\ref{Fig:LLnumer}~(b)-(c) is that LLs from monolayer-like sector move away from neutrality point as the bias is increased. This fact can also be understood using intuition developed about the band structure evolution under the bias: hybridization and floating away of monolayer-like bands [see Fig.~\ref{Fig:ABAband}~(a)-(b)] corresponds to monolayer-like LLs moving away and exhibiting avoided crossings with bilayer-like LLs. 

\begin{figure}[h]
\begin{center}
\setlength{\unitlength}{\columnwidth}
\begin{picture}(1,0.87)
\put(-0.01,0){\includegraphics[width=0.99\columnwidth]{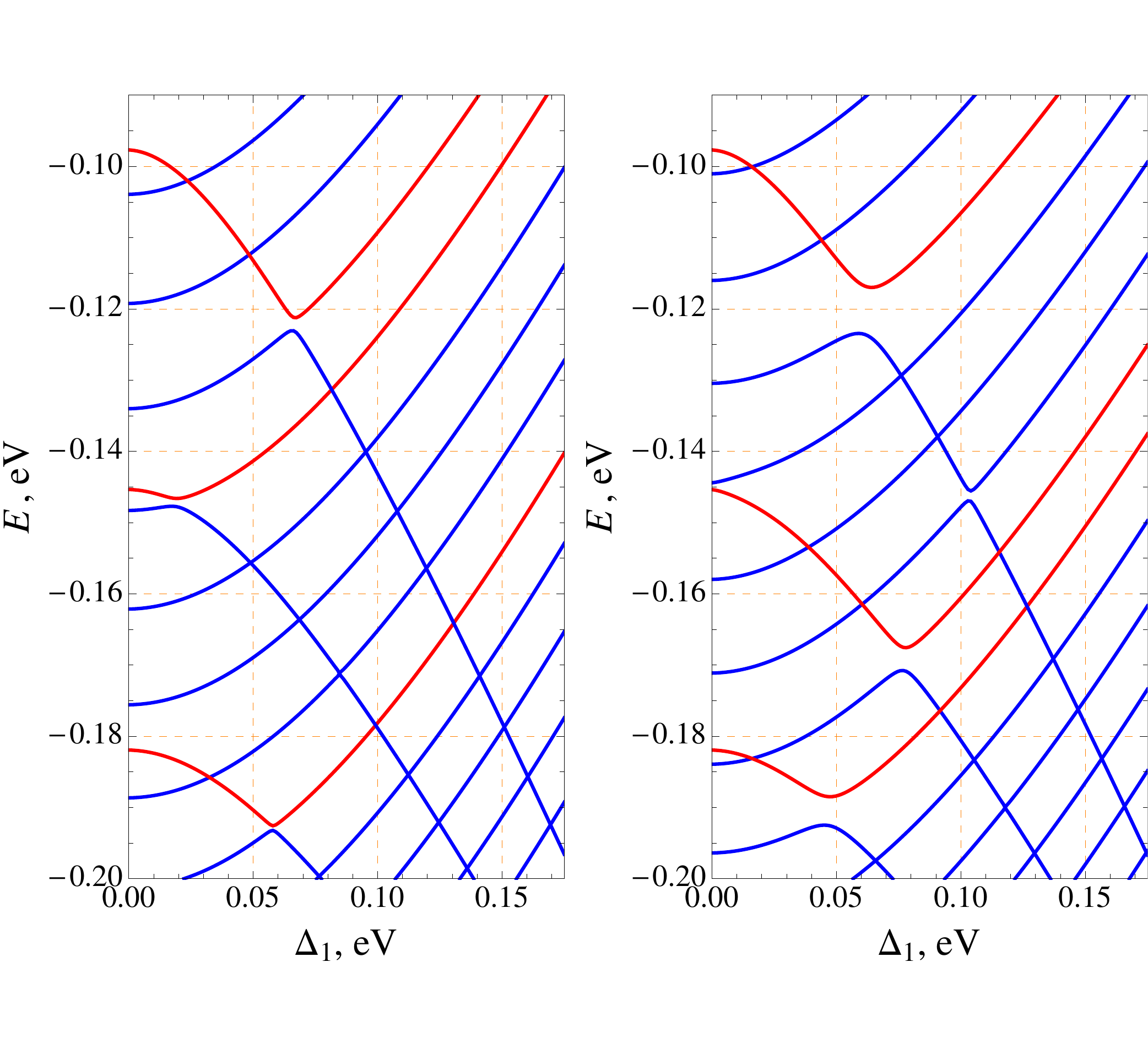}}
\put(0.2, 0.85){$\gamma_3 = 0.39$~\text{eV}}
\put(0.7, 0.85){$\gamma_3 = 0.78$~\text{eV}}
\put(0.05, .85){(a)}
\put(0.55, .85){(b)}
\put(0.055, 0.79){\red {\footnotesize $1_+$}}
\put(0.055, 0.74){\blue {\footnotesize $6_+$}}
\put(0.055, 0.67){\blue {\footnotesize $7_+$}}
\put(0.055, 0.55){\blue {\footnotesize $8_+$}}
\put(0.055, 0.49){\red {\footnotesize $2_+$}}
\put(0.055, 0.46){\blue {\footnotesize $9_+$}}
\put(0.04, 0.37){\blue {\footnotesize $10_+$}}
\put(0.04, 0.3){\blue {\footnotesize $11_+$}}
\put(0.055, 0.25){\red {\footnotesize $3_+$}}
\put(0.04, 0.21){\blue {\footnotesize $12_+$}}
\put(0.56, 0.79){\red {\footnotesize $1_+$}}
\put(0.56, 0.74){\blue {\footnotesize $6_+$}}
\put(0.56, 0.67){\blue {\footnotesize $7_+$}}
\put(0.56, 0.58){\blue {\footnotesize $8_+$}}
\put(0.56, 0.46){\red {\footnotesize $2_+$}}
\put(0.56, 0.49){\blue {\footnotesize $9_+$}}
\put(0.545, 0.43){\blue {\footnotesize $10_+$}}
\put(0.545, 0.33){\blue {\footnotesize $11_+$}}
\put(0.56, 0.25){\red {\footnotesize $3_+$}}
\put(0.545, 0.22){\blue {\footnotesize $12_+$}}
\end{picture}
\caption{ \label{Fig:LLg3} (Color online)
Influence of the value of $\gamma_3$ on the pattern of avoided crossings between monolayer and bilayer-like LL. Magnetic field is $B=10$~T. For clarity, only LLs in the $K_+$ valley are shown. Plot (a) is for $\gamma_3 = 0.39$~\text{eV} used in this paper, whereas plot (b) is for twice larger value of $\gamma_3$.  Labelling scheme of LLs is explained in the caption of Fig.~\ref{Fig:LLnumer}.
}
\end{center}
\end{figure}
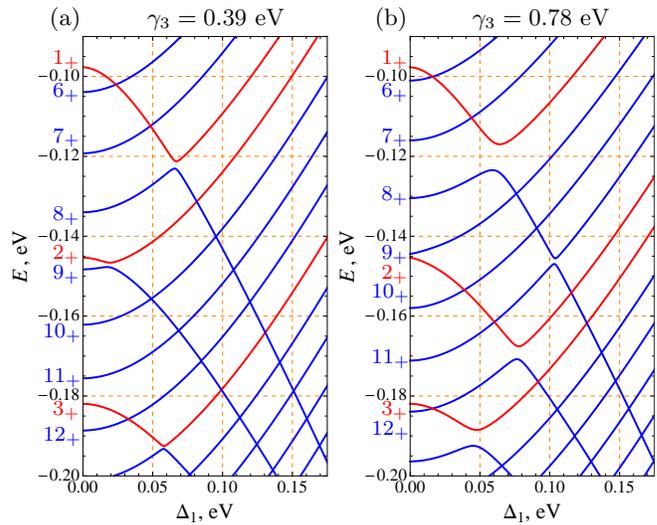

Interestingly, it is tight binding parameter $\gamma_3$ that controls the behavior of aforementioned avoided crossings. For example, let us zoom in into part of LLs spectrum corresponding to negative filling factors of order $\nu \approx -20$. We note that such filling factors are easily accessible in the experiment.~\cite{ThitiExp}  Fig.~\ref{Fig:LLg3} illustrates the pattern of avoided crossings for moderate bias $\D_1 = 0\div 0.17$~eV and two different values of $\gamma_3$: the one which is used in all simulations throughout this paper~[$\gamma_3=0.39$~eV, Fig.~\ref{Fig:LLg3}~(a)]  and a twice larger value~[$\gamma_3=0.78$~eV, Fig.~\ref{Fig:LLg3}~(b)]. 

\begin{figure*}
\begin{center}
\setlength{\unitlength}{\columnwidth}
\begin{picture}(2,.93)
\put(-.05,0){\includegraphics[width=0.66\columnwidth]{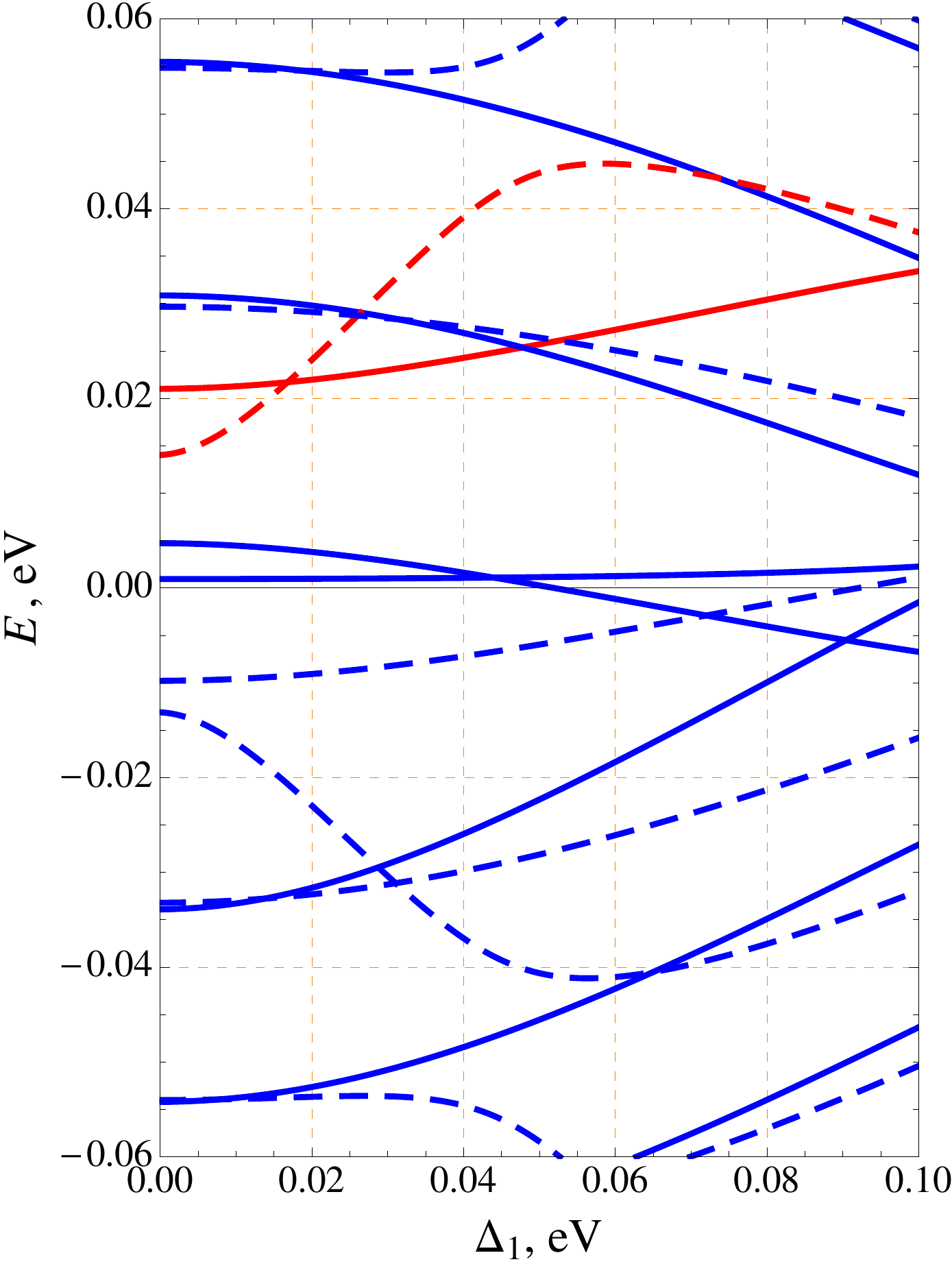}\quad
\includegraphics[width=0.66\columnwidth]{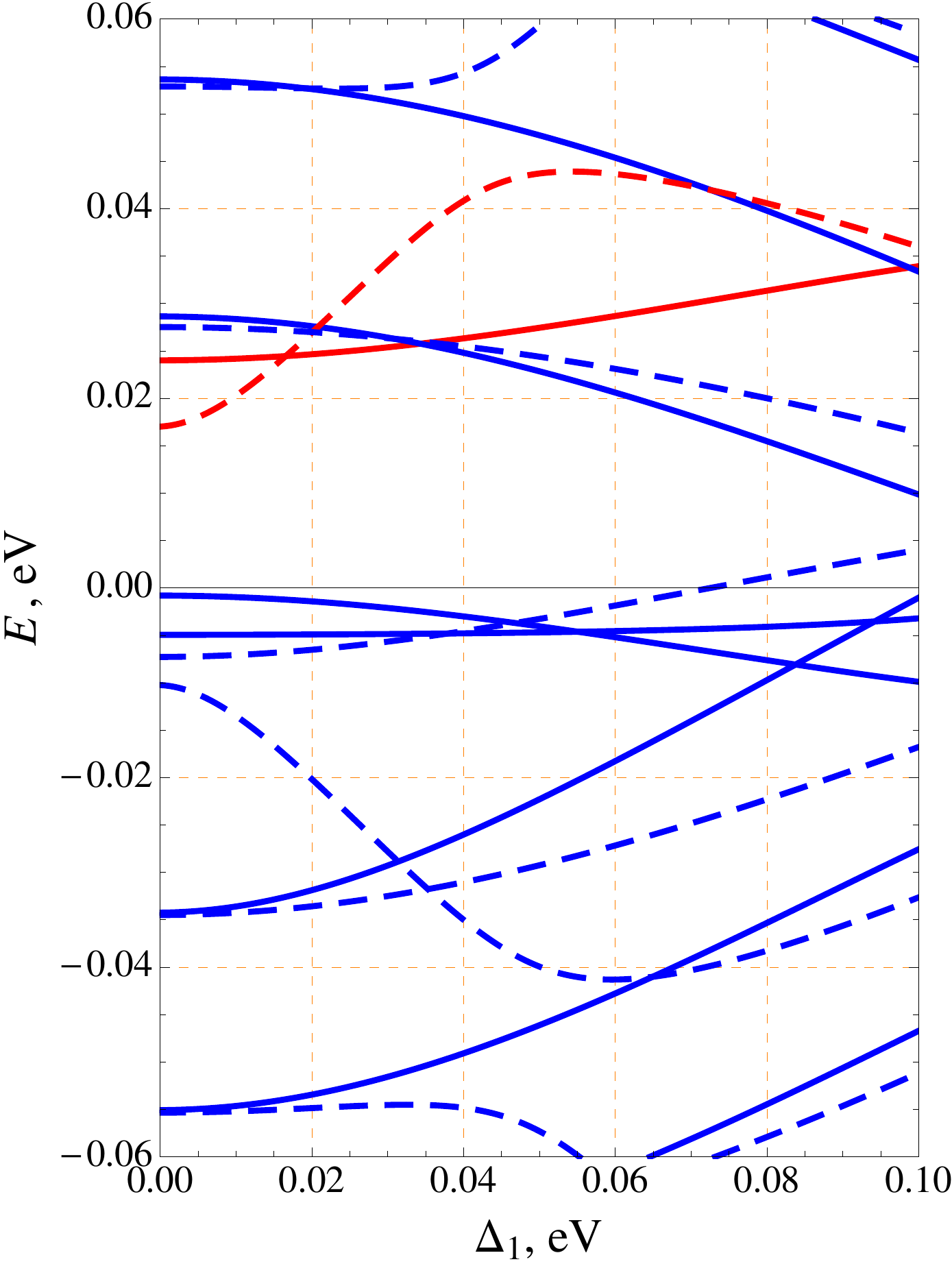}\quad
\includegraphics[width=0.66\columnwidth]{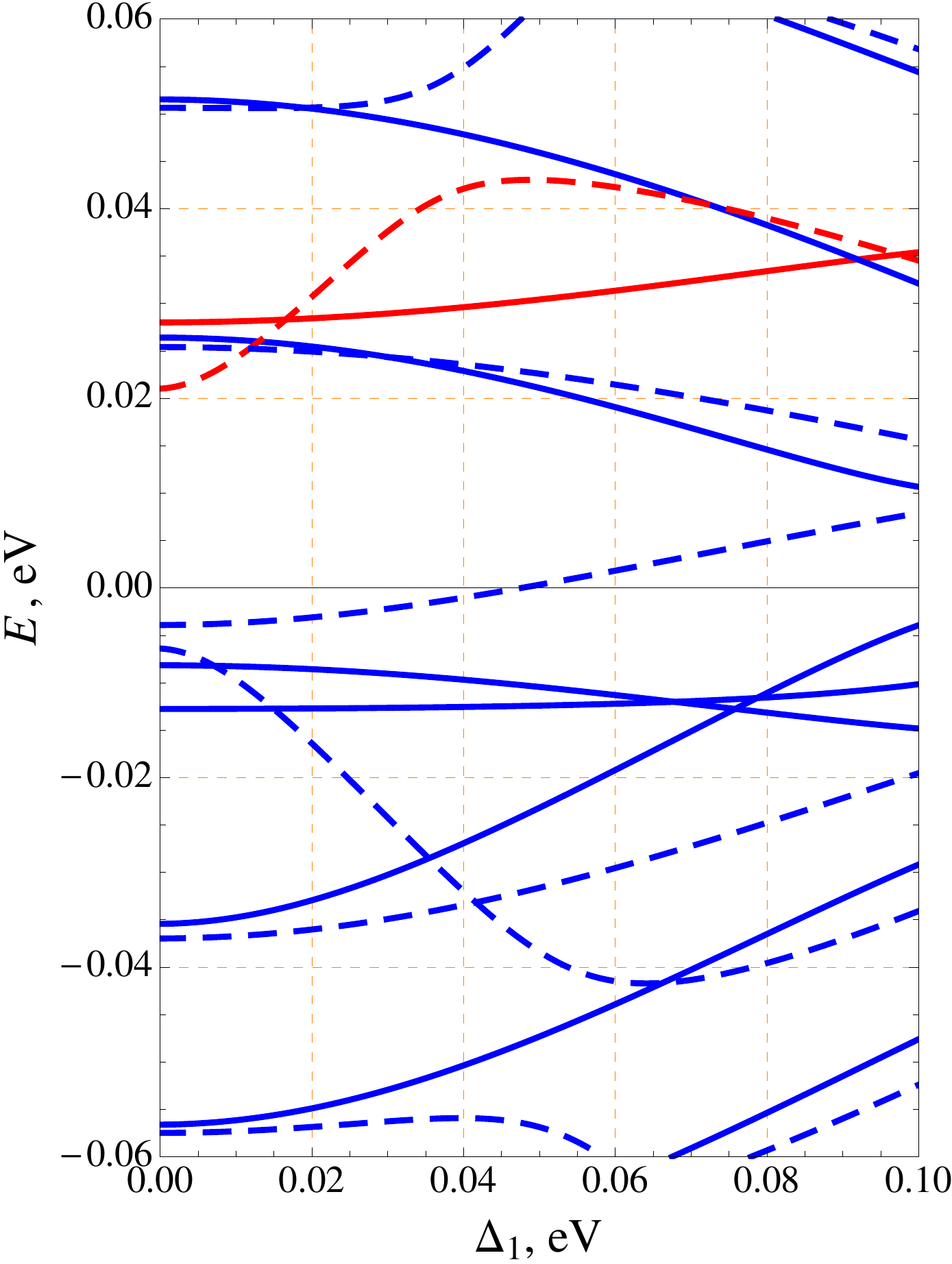}}
\put(0.05, .9){(a)}
\put(0.25, 0.9){$\Delta_2 = 0$}
\put(0.75, .9){(b)}
\put(0.92, 0.9){$\Delta_2 = 3$~\text{meV}}
\put(1.45, .9){(c)}
\put(1.62, 0.9){$\Delta_2 = 7$~\text{meV}}
\put(0.01, 0.82){\blue {\footnotesize $3_\pm$}}
\put(0.01, 0.66){\blue {\footnotesize $2_\pm$}}
\put(0.01, 0.62){\red {\footnotesize $0_+$}}
\put(0.01, 0.55){\red {\footnotesize $0_-$}}
\put(0.01, 0.51){\blue {\footnotesize $1_+$}}
\put(0.01, 0.485){\blue {\footnotesize $0_+$}}
\put(0.01, 0.4){\blue {\footnotesize $1_-$}}
\put(0.01, 0.37){\blue {\footnotesize $0_-$}}
\put(0.01, 0.25){\blue {\footnotesize $2_\pm$}}
\put(0.01, 0.1){\blue {\footnotesize $3_\pm$}}
\put(0.71, 0.82){\blue {\footnotesize $3_\pm$}}
\put(0.71, 0.66){\blue {\footnotesize $2_\pm$}}
\put(0.71, 0.63){\red {\footnotesize $0_+$}}
\put(0.71, 0.56){\red {\footnotesize $0_-$}}
\put(0.71, 0.48){\blue {\footnotesize $1_+$}}
\put(0.71, 0.435){\blue {\footnotesize $0_+$}}
\put(0.71, 0.405){\blue {\footnotesize $1_-$}}
\put(0.71, 0.375){\blue {\footnotesize $0_-$}}
\put(0.71, 0.24){\blue {\footnotesize $2_\pm$}}
\put(0.71, 0.1){\blue {\footnotesize $3_\pm$}}
\put(1.41, 0.8){\blue {\footnotesize $3_\pm$}}
\put(1.41, 0.63){\blue {\footnotesize $2_\pm$}}
\put(1.41, 0.66){\red {\footnotesize $0_+$}}
\put(1.41, 0.565){\red {\footnotesize $0_-$}}
\put(1.41, 0.39){\blue {\footnotesize $1_+$}}
\put(1.41, 0.36){\blue {\footnotesize $0_+$}}
\put(1.41, 0.44){\blue {\footnotesize $1_-$}}
\put(1.41, 0.415){\blue {\footnotesize $0_-$}}
\put(1.41, 0.23){\blue {\footnotesize $2_\pm$}}
\put(1.41, 0.095){\blue {\footnotesize $3_\pm$}}
\end{picture}
\caption{ \label{Fig:LLlarge} (Color online)
Landau levels as a function of displacement field for different values of~$\D_2$. Magnetic field $B=10$~T. Plot (a) uses zero value of $\D_2$, for plots (b) and (c) $\D_2=3$~meV, and $\D_2=7$~meV respectively. Value of $\D_2 = 7$~meV is sufficient to switch the order of $\blue (1,0)_-$ and $\blue (1,0)_+$ LLs from bilayer sector. }
\end{center}
\end{figure*}
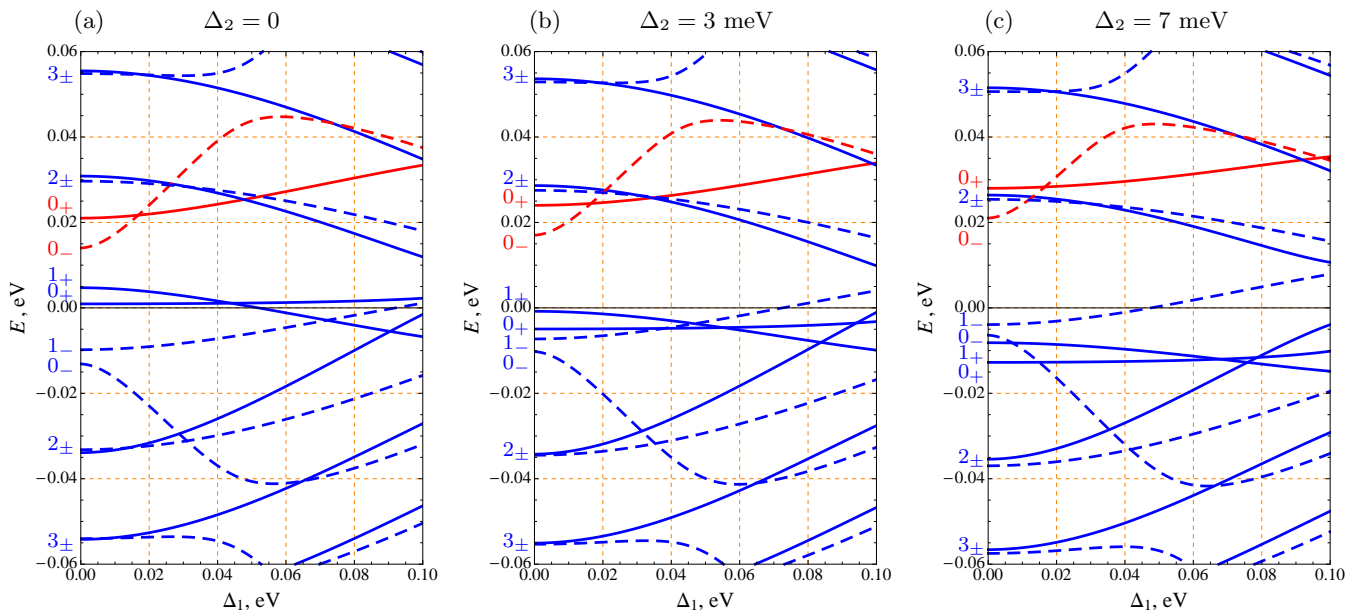

In Fig.~\ref{Fig:LLg3}(b), corresponding to the larger value of $\gamma_3$, the crossing pattern changes due to a slightly different order of LLs at zero bias. More noteworthy is the pronounced enhancement of gaps at anti-crossings. Due to this enhancement, one can identify the clear pattern in Fig.~\ref{Fig:LLg3}~(b): avoided crossings occur between a given LL originating from monolayer-like band and every \emph{third} bilayer-like LL. This pattern can be easily explained using the effect of trigonal warping on the LL wave functions. If, without a trigonal warping term, a given LL contains orbital $|i\rangle$ in its wave function, a non-zero trigonal warping admixes orbitals $|i\pm3\rangle, |i\pm6\rangle, \ldots$ with progressively smaller coefficients.  It is this admixture that underlies the avoided crossings, thus, it is natural that stronger trigonal warping enhances the gaps at anti-crossings. 

As an example, let us consider LL $1^\text{m}_+$ in the monolayer-like block~[labeling scheme of LLs is introduced in the caption of  Fig.~\ref{Fig:LLnumer}]. From Fig.~\ref{Fig:LLg3} we see that it exhibits a pronounced anti-crossing with $8^\text{b}_+$ bilayer-like LL, and also with $11^\text{b}_+$ LL for larger $\gamma_3$. In the absence of trigonal warping, the wave function of $1^\text{m}_+$ LL at $\D_1=0$ consists of a combination of $|0\rangle$  and $|1\rangle$ orbitals on $A_1-A_3$ and $B_1-B_3$ sublattices respectively~[see Eq.~(\ref{e:basis2})]. The wave function of $8^\text{b}_+$ LL is dominated by   $|6\rangle$ and $|8\rangle$ orbitals on symmetric combination of $A_{1,3}$ sublattices and $B_2$ sublattice. Non-zero trigonal warping changes the wave functions, admixing substantial components of  $|0\rangle, |3\rangle, |9\rangle,\ldots$ orbitals to the $|6\rangle$ orbital on $A_{1}+A_{3}$ sublattice. Finally, non-zero $\D_1$ breaks inversion symmetry, and initially orthogonal symmetric and antisymmetric combination of $A_{1,3}$ sublattices evolve to have non-vanishing overlap. This gives rise to the anti-crossings observed in Fig.~\ref{Fig:LLg3}.

The observed sensitivity of the pattern of avoided crossings can be potentially used to fix the value of the tight-binding parameter $\gamma_3$. However, in order to do this, the other tight-binding parameters have to be refined. Below we discuss how information about them can be extracted from the LLs crossings at small filling factors and a smaller bias.

\subsection{Pattern of Landau level crossings at smaller bias and its sensitivity to tight-binding parameters}
Having discussed the behavior of LL at high bias and (or) large filling factors, we concentrate on the regime of small bias, when the new Dirac points have not formed yet. To better understand the complicated pattern of LL crossings that arises in this regime, we zoom in on the region $\Delta_1\leq 0.1 {\rm eV}$, and concentrate only on several low-lying LLs [see Fig.~\ref{Fig:LLlarge}~(a)]. 

Interestingly, the bias {\it completely} lifts the degeneracies of all LLs. Therefore, we expect that in biased bilayer graphene all quantum Hall plateaus will be spaced by $e^2/h$. Notice that this goes against intuition acquired from the studies of monolayer and bilayer graphene,~\cite{CastroNeto09} where in the absence of interactions Landau levels remain valley-degenerate. 

Landau levels undergo a series of crossings, as illustrated in Fig.~\ref{Fig:LLlarge}(a). In particular, level $0^\text{b}_-$ moves down, crossing weakly split $2^\text{b}_\pm$ LLs at $\Delta_1\approx 0.03 \, {\rm eV}$. The energy of level $1^\text{b}_-$ increases with bias, such that it crosses level $1^\text{b}_+$ at $\Delta_1\approx 0.07\, {\rm eV}$. Furthermore, levels $0^\text{b}_+$ and $1^\text{b}_+$ cross at $\Delta_1\approx 0.05\, {\rm eV}$ as $1^\text{b}_+$ level slowly moves down, and $0^\text{b}_+$ level is essentially non-dispersive. Finally, monolayer-like levels $0^\text{m}_-$ and $0^\text{m}_+$ both float up and cross at small bias $\Delta_1\approx 0.015\, {\rm eV}$. They also cross $2^\text{b}_{\pm}$ LLs which move down. 

As we emphasized above, although qualitatively the band structure of trilayer graphene is rather well understood at this point, not all parameters in the tight-binding model are precisely known. The complex LL crossing pattern described above is very sensitive to the choice of tight-binding parameters, and may provide a valuable tool for fixing their values. We illustrate this by exploring how LL crossing pattern changes when parameter $\Delta_2$ is varied. Such a parameter  is allowed by symmetry, but have not been taken into account previously. 

The LL crossing pattern for $\Delta_2=3\, {\rm meV}$ and $\Delta_2=7 \, {\rm meV}$ is illustrated in Fig.~\ref{Fig:LLlarge}(b)-(c). It is evident that the crossing between $1^\text{b}_-$ and $1^\text{b}_+$ is significantly shifted even by small $\Delta_2=3\, {\rm meV}$; so does the crossing between monolayer-like $0^\text{m}_+$ LL and $2^\text{b}_{\pm}$ levels.  

As $\Delta_2$ is increased further [see Fig.~\ref{Fig:LLlarge}~(c)], new crossings appear: in particular, a pair $0^\text{b}_+,1^\text{b}_+$ swaps positions with the pair $0^\text{b}_-,1^\text{b}_-$ at zero bias [as expected from Eq.~(\ref{Eq:0LLBLG})]. This leads to new crossings between $0^\text{b}_-$ and $0^\text{b}_+, 1^\text{b}_+$ levels at a small bias, $\Delta_1\approx 0.01\, {\rm eV}$. Also, $0^\text{m}_+$ LL moves above $2_\pm$ levels, such that crossings between them disappear. We see that even small variations of parameter $\Delta_2$ can change the pattern of LL crossings. Similarly, the positions of different crossings are sensitive to the values of $\gamma_2, \gamma_5, \delta$, although in a more subtle manner. 

As we pointed out, non-zero $\D_2$ can be present among other tight-binding parameters. Moreove, if one accounts for the effect of electric field self-consistently, in addition to renormalization of  $\D_1$~(corresponding to screening), parameter $\D_2$ changes as well. We found that typical values of $\D_2$ obtained from such self-consistent procedure is $2\, {\rm meV}$ at charge neutrality point.~\cite{tobe} We also performed numerical simulations for the case when magnetic field is present~(detailed procedure will be described elsewhere). We found that the self-consistent value of $\D_2$ in the presence of magnetic field can be enhanced compared to the case when $B=0$, due to discreteness of LLs. For example, from Eqs.~(\ref{Eq:0LLBLGK+})-(\ref{Eq:1LLBLGK+}) one can see that at zero bias wave functions of bilayer-like $0^\text{b}_+$ and $1^\text{b}_+$ LLs are localized on the middle layer. Consequently, by emptying these LLs one can create an excess positive charge on the middle layer, thus generating positive non-zero $\D_2$ that can be as large as $4\, {\rm meV}$.~\cite{tobe}

\section{Discussion and Outlook \label{Sec:Discussion}}

In summary, we have studied the electronic properties of the $ABA$-stacked trilayer graphene in the perpendicular electric field. We found that the hybridization of the monolayer-like and bilayer-like bands leads to an opening of a small ($\lesssim 10\, {\rm meV}$) band gap at experimentally achievable values of bias $\Delta_1\lesssim 0.25\, {\rm eV}$. At bias $\Delta_1\gtrsim 0.1\, {\rm eV}$, a new set of Dirac points appears in each valley, which includes one central DP, and six off-center DPs. Masses of the emergent Dirac fermions are tunable by bias, and, interestingly, some of the masses vanish at a bias which for our choice of tight-binding parameters corresponds to $\Delta_1^*\approx 0.185 \, {\rm eV}$. Therefore, the band gap depends on the bias in a non-monotonic manner and closes at the bias $\Delta_1^*$. This behavior should be contrasted with bilayer graphene,~\cite{Castro07,Oostinga08,Zhang09} where band gap grows monotonically as a function of bias, and can become very large, of the order of $0.25\, {\rm eV}$.~\cite{Zhang09}

We have also studied the spectrum of LLs, finding that at bias $\Delta_1\gtrsim 0.1\, {\rm eV}$ the band structure transformation gives rise to new three-fold-degenerate groups of LLs. The evolution of low-lying LLs from the unbiased case, where the spectrum consists of LLs in the monolayer and bilayer sectors, to the degenerate groups of LLs at higher bias, is accompanied by multiple level crossings. 
Experimentally, these LL crossings will give rise to phase transitions between quantum Hall states as a function of bias. We have also shown that the pattern of crossings depends strongly on the values of certain tight-binding parameters, such as $\Delta_2$ and $\gamma_3$. Studying the phase transitions in QHE regime experimentally should enable the precise determination of these parameters. 

Our results show that single-particle splittings between LLs in biased trilayer are sizable, of the order of a few meV, and exceed the Zeeman energy of $1\, {\rm meV}$ at typical magnetic field of $B=10\, {\rm T}$. This is expected to play an important role in the analysis of quantum Hall ferromagnetism in trilayer graphene,~\cite{GelderenFerroTrilayer,MacDonald-QHF3,Yoshioka-skyrme} which will be a subject of subsequent publication.~\cite{tobe} Here we just note that we found that, owing to the small LL spacing, the effective Coulomb interactions are strongly screened,~\cite{tobe} similar to the case of bilayer graphene~\cite{Gorbar} and, as a result, the Coulomb energy scale becomes comparable to typical single-particle energy splitting. This situation is quite different from the case of monolayer graphene, where single-particle splittings of LLs are much smaller than the interaction energy.~\cite{Abanin06,Nomura06,Goerbig06,Young12} 

Our study reveals the unique tunability of Landau levels in trilayer graphene: bias controls their energies, wave functions, and degeneracies. We speculate that this may enable the exploration of interesting interaction-induced phenomena in the QHE regime. First, the three-fold degeneracy of LLs at higher bias may allow one to study quantum Hall ferromagnetism, skyrmions, and fractional quantum Hall states with $SU(3)$-symmetry. Second, the bands in trilayer graphene are strongly trigonally warped, which provides an opportunity to study the effect of anisotropic band structure on quantum Hall ferromagnetism and fractional QHE.~\cite{Papic_anis} Third, Landau level wave functions involve a superposition of different non-relativistic LL orbitals with weights tunable by bias. Therefore the LL form-factors are tunable, which may allow one to realize new effective interaction regimes, stabilize desired fractional quantum Hall states and drive phase transitions between them.~\cite{Papic11,Papic12}

We believe that phenomena discussed above can be experimentally observed in the near future. We note that the observation of effects associated with the emergent DPs is possible when disorder broadening is smaller than the energy scale over which effective description in terms of DPs is valid. Fig.~\ref{Fig:ABAband} indicates that the latter energy scale is of the order of $20\,{\rm meV}$, which is much larger than the disorder broadening achievable in trilayer graphene on hexagonal boron nitride.~\cite{ThitiExp,Campos12} Therefore, new DPs should be observable in currently available samples. 

Finally, we note that very recently multiple phase transitions between different quantum Hall states in biased trilayer graphene have been reported.~\cite{Lau12} It is possible that these phase transitions can be understood in terms of single-particle LL spectrum evolution described in our paper.

\section*{Acknowledgements}
We thank Pablo Jarillo-Herrero, Leonardo Campos, and Thiti Taychatanapat for attracting our attention to the problem of biased trilayer graphene, and for many helpful discussions.

\emph{Note added.}---When this manuscript was at the final stages of preparation, we became aware of a recent work by Morimoto and Koshino, Ref.~\onlinecite{Koshino12}, where new bias-induced Dirac points in trilayer graphene were also predicted.

\end{document}